\documentclass[twocolumn,showpacs,preprintnumbers,amsmath,amssymb,prb,superscriptaddress]{revtex4-1}
\pdfoutput=1
\usepackage{graphicx}
\usepackage{dcolumn}
\usepackage{bbold}
\usepackage{color}

\usepackage[low-sup]{subdepth}
\usepackage{multirow}
\usepackage{amsfonts}
\usepackage[english]{babel}
\usepackage[T1]{fontenc}
\usepackage{times}
\usepackage[colorlinks,bookmarks=true,citecolor=blue,linkcolor=red,urlcolor=blue]{hyperref}
\usepackage[tight, FIGTOPCAP, hang, raggedright, nooneline]{subfigure}

\usepackage{verbatim}
\usepackage{bm}
\usepackage{mathrsfs}

\usepackage{float}

\newcommand* {\vek}[1]{{\ensuremath{\bm{\mathrm{#1}}}}}

\newcommand* {\bra}[1]{\ensuremath{\langle {#1} |}}
\newcommand* {\ket}[1]{\ensuremath{| {#1} \rangle}}
\newcommand* {\braket}[1]{\ensuremath{\langle {#1} \rangle}}

\usepackage{amsmath,amssymb}
\usepackage{color}
\usepackage{graphicx}
\usepackage{natbib}

\def\zt{\mathbb{Z}_2}
\def\zth{\mathbb{Z}_3}
\def\zn{\mathbb{Z}_n}
\def\e{\epsilon}
\def\D{\Delta}
\def\eff{\text{eff}}

\begin{document}

\title{Nonuniform Parafermion Chains: Low-Energy Physics and Finite-Size Effects}

\author{Mohammad Mahdi Nasiri Fatmehsari}
\affiliation{Pasargad Institute for Advanced  Innovative Solutions (PIAIS), Tehran 19395-5531, Iran}
\affiliation{Department of Converging Technologies, Khatam University, Tehran 19395-5531, Iran}

\author{Mohammad-Sadegh Vaezi}
\email{Corresponding Author: sadegh.vaezi@piais.ac.ir, sadegh.vaezi@gmail.com}
\affiliation{Pasargad Institute for Advanced  Innovative Solutions (PIAIS), Tehran 19395-5531, Iran}
\affiliation{Department of Converging Technologies, Khatam University, Tehran 19395-5531, Iran}

\date{\today}

\begin{abstract}

The nonuniform $\zt$ symmetric Kitaev chain, comprising alternating topological and normal regions, hosts localized states known as edge-zero modes (EZMs) at its interfaces. These EZMs can pair to form qubits that are resilient to quantum decoherence, a feature expected to extend to higher symmetric chains, i.e., parafermion chains. However, finite-size effects may impact this ideal picture. Diagnosing these effects requires first a thorough understanding of the low-energy physics where EZMs may emerge. Previous studies have largely focused on uniform chains, with nonuniform cases inferred from these results. While recent work [Narozhny, Sci. Rep. 7, 1447 (2017)] provides an insightful analytical solution for a nonuniform $\zt$ chain with two topological regions separated by a normal one, its complexity limits its applicability to chains with more regions or higher symmetries. Here, we present a new approach based on decimating the highest-energy terms, facilitating the scalable analysis of $\zn$ chains with any number of regions. We provide analytical results for both $\zt$ and $\zth$ chains, supported by numerical findings, and identify the critical lengths necessary to preserve well-separated EZMs.

\end{abstract}

\maketitle

\section{Introduction}

\begin{figure}
    \centering 
    \centerline{\includegraphics[width=1.05\linewidth]{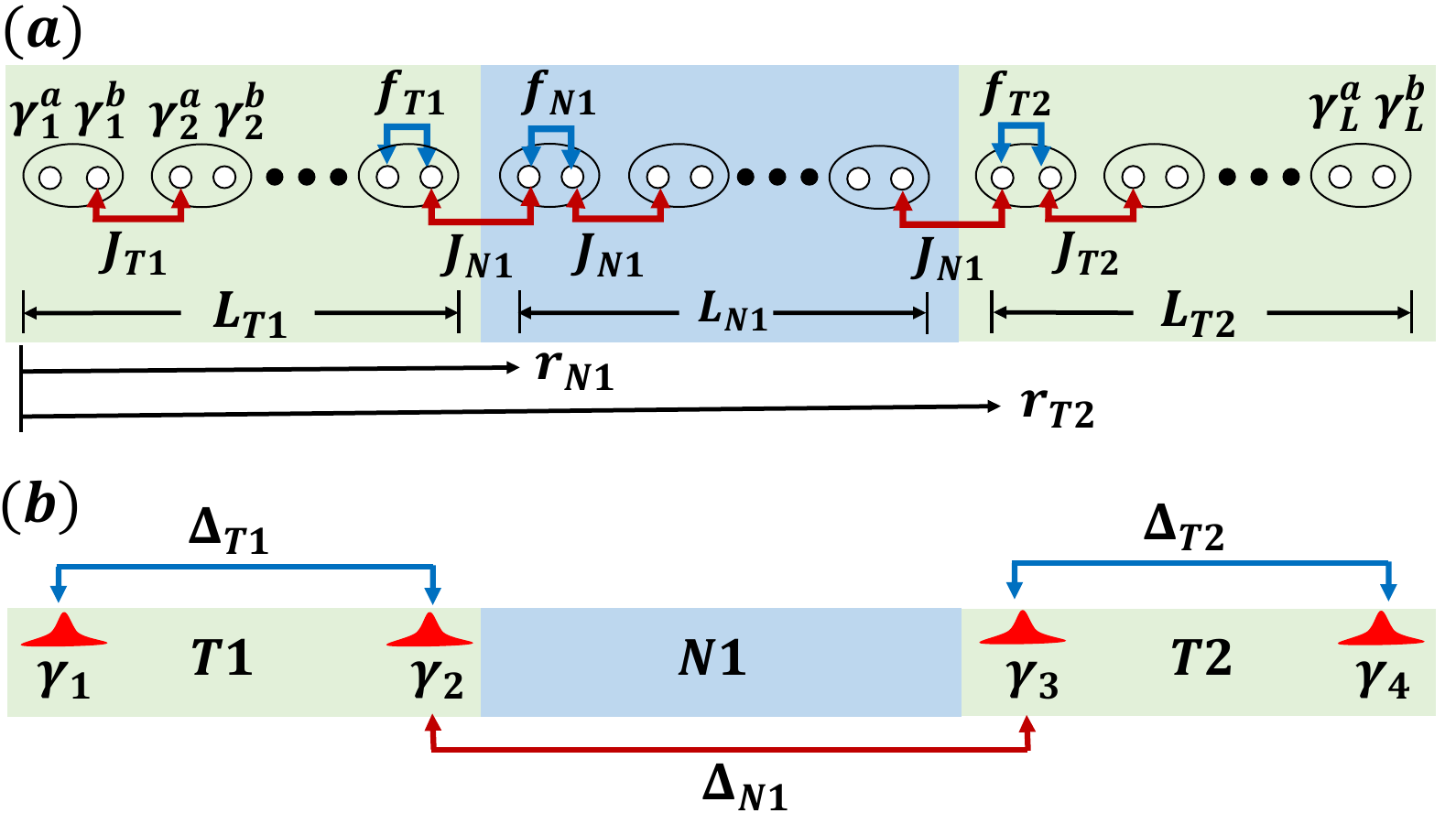}}
    \caption{(a)  Schematic of a $\zt$ symmetric chain featuring  two topological regions (T1, T2) separated by a normal one (N1). Green and blue areas refer to T and N regions, respectively. For $\zth$, substitute $\gamma$'s with $\eta$'s.   (b) Four expected EZMs are highlighted in red.}
    \label{NT2} 
\end{figure}

\indent Quantum computation has long grappled with the challenge of quantum decoherence~\cite{nielsen2010quantum, rieffel2011quantum, divincenzo1999quantum, knill1997theory, lidar1998decoherence, zurek1991decoherence, schlosshauer2019quantum, chuang1995quantum, shor1996fault, shor1995scheme}.
Among various efforts, it has been demonstrated that topological systems such as Kitaev chain with $\zt$ symmetry~\cite{kitaev2001unpaired, leumer2020exact, ranvcic2022exactly, sung2023simulating, miao2018majorana, burnell2013measuring} and its generalizations, i.e., parafermion chains with $\zn$ ($n\geq3$) symmetry~\cite{fendley2012parafermionic, fendley2014free, alicea2016topological, jermyn2014stability, iemini2017topological, vaezi2013fractional,  vaezi2017numerical, mahyaeh2020study, zhuang2015phase, wu2002qubits}, can help tackle this issue (for convenience, we will henceforth refer to the Kitaev chain as the $\zt$-symmetric parafermion chain).
These models exhibit distinct phases—gapped normal (N) and topological (T) separated by a critical point. 
Notably, within the T phase, certain decoherence-protected degrees of freedom manifest at the edges known as edge zero modes (EZMs). These EZMs, which commute with the Hamiltonian and display non-Abelian statistics~\cite{nayak2008non,rao2017introduction}, are resilient to local perturbations, making them suitable for constructing robust qudits (for $\zt$ symmetry, the term 'qubit' is commonly used).
However, the need for quantum computation goes beyond a single qudit.
To reach this goal we may use many copies of these chains in T phase, e.g., $n_T$, while making sure that the EZMs from the end of different regions are far enough that cannot hybridize and disappear.
The common practice is linking $n_T$ T-regions via $n_N$ ($= n_T - 1$) N-regions (Fig.~\ref{NT2}) which act as isolaters.
This would result in $2 \times n_T$ EZMs at the boundaries and a $n^{n_T}$-fold topological ground sate degeneracy.

\indent We know that the above picture is idealized, as in reality, due to finite-size effects, there would be some splitting among the $n^{n_T}$ states. If the splitting is significant, it may cause hybridization and dephasing, making the quantum computation less fault-tolerant. Analytical results studying the finite-size effects on the low-energy states of such systems, where EZMs emerge and play a key role in quantum computation, are rare, especially for $\zth$ and higher-symmetry parafermion chains with $N_T > 1$.
This scarcity is largely due to the challenges of directly deriving the low-energy Hamiltonian of nonuniform chains. In Ref.~\cite{narozhny2017majorana}, an analytical study of the low-energy behavior for a $\zt$ chain with $n_T = 2$ was presented; however, the approach is complex and becomes impractically challenging for cases with $n_T > 2$ or for chains with $\zn$ symmetry. Thus, a new method that simplifies this process would be highly valuable for advancing studies in this area.

\indent In this work, we show that by implementing the Decimation of the Highest-Energy term (DHE) method~\cite{dasgupta1980low,fisher1995critical,shivamoggi2011majorana,benhemou2023universality}, one not only recovers the result of~\cite{narozhny2017majorana} but also extends it to cases with $n_T \geq 3$. This method also enhances our understanding of parafermion chains with arbitrary $n_T$, particularly for $\zth$ symmetry, an area less explored compared to $n_T = 1$. Our findings highlight the versatility of this approach across various symmetries. To validate our results using the DHE method, we employ numerical techniques such as Exact Diagonalization and Density Matrix Renormalization Group (DMRG)~\cite{white1992density, white1993density, schollwock2005density, schollwock2011density, verstraete2023density, vaezi2017entanglement}. Finally, our analysis of the ground state energy splitting underscores the importance of region lengths in ensuring well-separated EZMs for optimal performance.

\indent The paper is organized as follows: Section I covers the nonuniform $\zt$ model, followed by the $\zth$ model in Section II. Section III addresses ground state energy splitting from finite-size effects, and Section IV provides concluding remarks.

\section{$\zt$ Symmetry}
The Hamiltonian of a $\zt$ chain with $n_T$ number of T regions reads
\begin{eqnarray}
\label{H_Z2_main}
H_{\zt} = \sum^{n_T}_{i = 1} H_{Ti} + \sum^{n_N}_{i = 1} H_{Ni},
\end{eqnarray}
where
\begin{eqnarray}
\label{H_Z2_TR}
H_{Ti} &=& 
i J_{Ti}\sum^{\vek{r}_{Ti} + L_{Ti}-1}_{j = \vek{r}_{Ti} +1} {\gamma^{b}_{j} \gamma^{a}_{j+1}} 
+ if_{Ti} \sum^{\vek{r}_{Ti} + L_{Ti}}_{j = \vek{r}_{Ti}+1} {\gamma^{a}_{j} \gamma^{b}_{j}} ,\nonumber\\
H_{Ni} &=& 
i J_{Ni}\sum^{\vek{r}_{Ni} + L_{Ni}}_{j = \vek{r}_{Ni}} {\gamma^{b}_{j} \gamma^{a}_{j+1}}
+ if_{Ni} \sum^{\vek{r}_{Ni} + L_{Ni}}_{j = \vek{r}_{Ni}+1} {\gamma^{a}_{j} \gamma^{b}_{j}}.
\end{eqnarray}
Here, $\gamma$'s are the Majorana fermion operators satisfying the relations $\gamma = \gamma^{\dag}$, and $\{\gamma^{\alpha}_i,\gamma^{\beta}_j\} = 2 \delta_{ij}\delta_{\alpha \beta}$.
The non-negative couplings of the $n$-th T (N) region with $L_{Ti}$ ($L_{Ni}$) number of sites satisfy the relation $f_{Ti}/J_{Ti} < 1$ ($f_{Ni}/J_{Ni} > 1$). 
For convenience, we have absorbed the interaction terms between T and N regions into the N regions.
The vectors $\vek{r}_{Ti}$ and $\vek{r}_{Ni}$ also refer to real space starting point of $i$-th T and N regions (see Fig.~\ref{NT2} (a)), respectively.

\begin{figure}[t]
    \centering 
    \centerline{\includegraphics[width=0.8\linewidth]{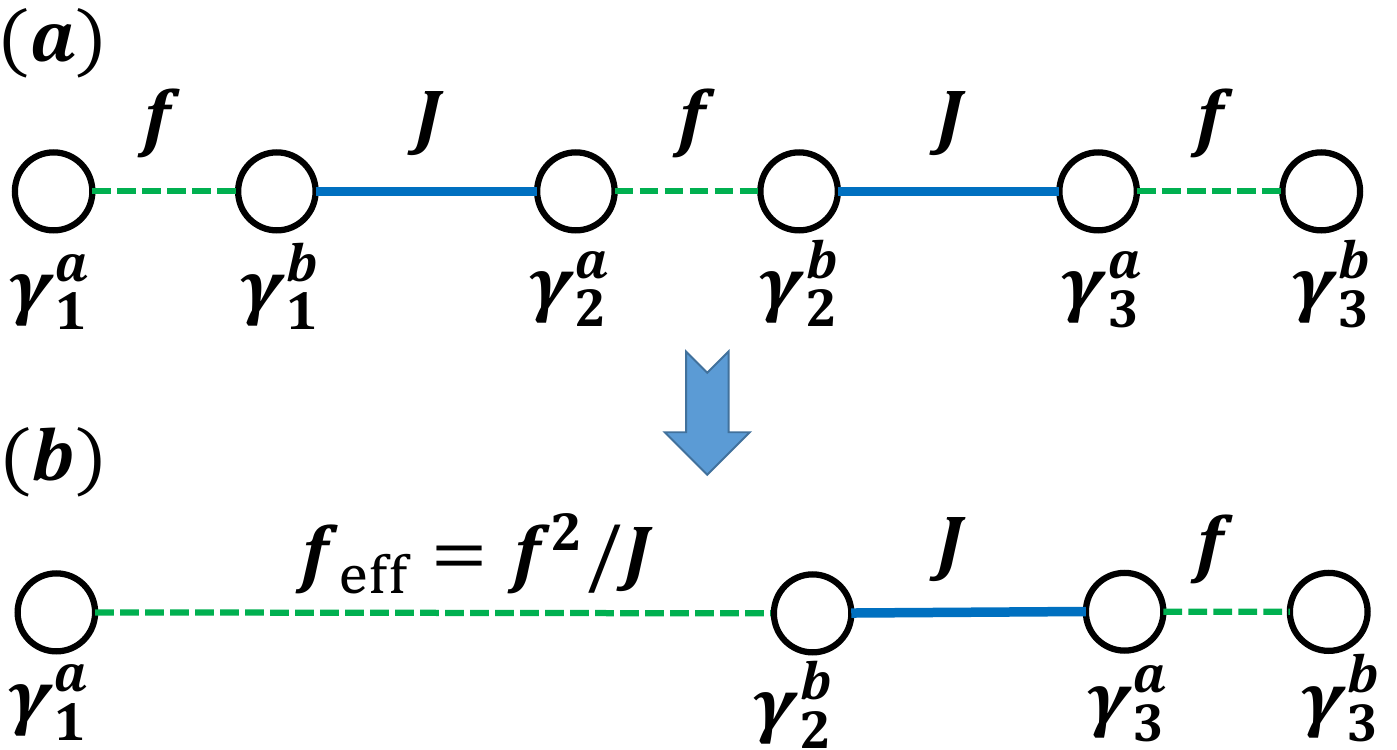}}
    \caption{Schematic illustration of the DHE method. Decimating the $J$ couplings $\gamma^{b}_{1}$ and $\gamma^{a}_{2}$ in panel (a) leads to the configuration of (b).}
    \label{DHE_z2} 
\end{figure}

\indent As stated earlier we aim to study low-energy physics where EZMs appear and hold particular significance for quantum computation by employing the DHE method.  
This method is an iterative process 
that involves eliminating the most energetic term within the Hamiltonian and replacing it with effective longer-range interactions using second-order perturbation theory at each iteration. For instance, within the T region where $J > f$, decimating $J$—let's say between $\gamma^{b}_{1}$ and $\gamma^{a}_{2}$ —leads to an effective interaction $f_{\eff} = \frac{f^2}{J}$ between $\tilde{\gamma}^{a}_{1}$ and $\tilde{\gamma}^{b}_{2}$, which can be approximated by $\gamma^{a}_{1}$ and $\gamma^{b}_{2}$, respectively (Fig.~\ref{DHE_z2}). Smaller $f/J$ ratios yield better approximations. Conversely, within the N region where $f > J$, one should decimate $f$ terms, resulting in an effective interaction $J_{\eff} = \frac{J^2}{f}$.
The details of this method applied to a single T and N regions are available in Ref.~\cite{shivamoggi2011majorana}.
Applying this method to Eq.~(\ref{H_Z2_TR}) and neglecting constant terms yields
\begin{eqnarray}
\label{H_Z2_TR_dec}
H_{\zt, Ti} &\approx&
i \D_{Ti} \gamma^{a}_{\vek{r}_{Ti} +1} \gamma^{b}_{\vek{r}_{Ti}+L_{Ti}}, \nonumber\\
H_{\zt, Ni} &\approx&
i \D_{Ni} \gamma^{b}_{\vek{r}_{Ni}} \gamma^{a}_{\vek{r}_{Ni}+L_{Ni}+1},
\end{eqnarray}
where
\begin{eqnarray}
\label{D_Z2}
\D_{Ti} = f_{Ti}(\frac{f_{Ti}}{J_{Ti}})^{L_{Ti-1}},~
\D_{Ni} = J_{Ni}(\frac{J_{Ni}}{f_{Ni}})^{L_{Ni}}.
\end{eqnarray}
Labeling the $\gamma^{a}$'s and $\gamma^{b}$'s in Eq.~(\ref{H_Z2_TR_dec}) according to their positions at the interfaces (Fig.~\ref{NT2}(b)) provides
\begin{eqnarray}
\label{H_Z2_eff}
H^{\eff}_{\zt} \approx
i \sum^{n_T}_{i=1} \D_{Ti} \gamma_{2i-1} \gamma_{2i} +
i \sum^{n_N}_{i=1} \D_{Ni} \gamma_{2i} \gamma_{2i+1}.
\end{eqnarray}
Interestingly, this outcome resembles the famous Kitaev chain.
One can easily confirm that for $n_T = 1$, the result for a single-region Kitaev chain~\cite{kitaev2001unpaired} is retrieved, i.e., $H^{\eff}_{\zt} = i \e_{1} \gamma_{1} \gamma_{2}$, where $\e_{1} = \D_{T1}$ is the ground state energy splitting.
The associated non-local quasi-fermion operators then reads $c_{12} \approx (\gamma_{1} + i \gamma_{2})/\sqrt{2}$.

\indent Unfortunately, for $n_T > 1$, the situation is not as trivial as in the single-region case. This becomes evident from Eq.(\ref{H_Z2_eff}), where the competition among $\D$'s plays a major role in determining which EZMs couple to form a quasi-fermion and which energy levels correspond to the lowest energies. However, as we will demonstrate, applying the DHE method to Eq.(\ref{H_Z2_eff}) can still be beneficial. To better understand this, let us first review the case $n_T = 2$, for which an analytical result is available~\cite{narozhny2017majorana}. Here, different scenarios are possible.
Consider, for example, two extreme cases:
\begin{center}
(a) ~$\D_{T1}, \D_{T2} \gg \D_{N1}$.\\
(b) ~$\D_{T1}, \D_{T2} \ll \D_{N1}$.\\
\end{center}
In case (a), if we further assume that $\D_{T1} < \D_{T2}$, then using the DHE method, we conclude that $\gamma_3$ couples with $\gamma_4$ with a coupling $\e_{34} = \D_{T2}$. By eliminating this pair (as part of the DHE procedure), we are left with $\gamma_1$ and $\gamma_2$, which are now compelled to couple with $\e_{12} = \D_{T1}$. That is, $H^{(a)}_{\zt,n_T=2} = i (\e_{12} \gamma_{1} \gamma_{2} + \e_{34} \gamma_{3} \gamma_{4})$.
The quasi-fermions are $c_{12} = (\gamma_1 + i \gamma_2)/\sqrt{2}$ and $c_{34} = (\gamma_3 + i \gamma_4)/\sqrt{2}$, with the corresponding low-energy states (LES) given by $|n_{12}, n_{34} \rangle$. Evidently, $\e_{12} < \e_{34}$. Note that if we had assumed $\D_{T1} > \D_{T2}$, then we would have instead $\e_{12} > \e_{34}$.
Case (b) is more interesting. Here, $\gamma_2$ and $\gamma_3$ are coupled first with $\e_{23} = \D_{N1}$. Continuing the DHE procedure and eliminating this pair, we are left with $\gamma_1$ and $\gamma_4$, which now interact with an effective strength $\D_{\eff} = \frac{\D_{T1} \D_{T2}}{\D_{N1}}$. Therefore, they couple with $\e_{14} = \D_{\eff}$, leading to $H^{(b)}_{\zt,n_T=2} = i (\e_{14} \gamma_{1} \gamma_{4} + \e_{23} \gamma_{2} \gamma_{3})$.
 The quasi-fermions are $c_{14} = (\gamma_1 + i \gamma_4)/\sqrt{2}$ and $c_{23} = (\gamma_2 + i \gamma_3)/\sqrt{2}$, with the corresponding LES given by $|n_{14}, n_{23} \rangle$. These results are in exact agreement with Ref.~\cite{narozhny2017majorana}.

\begin{figure}[t]
    \centering 
    \centerline{\includegraphics[width=0.65\linewidth]{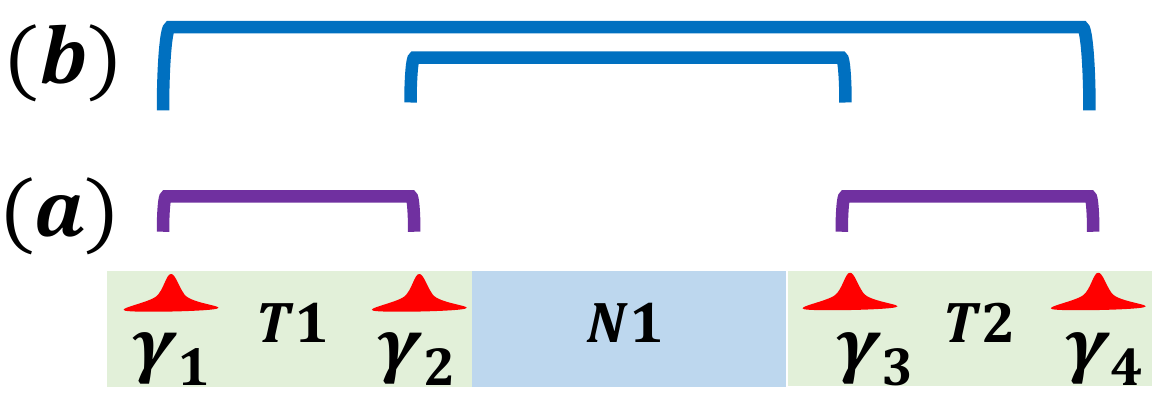}}
    \caption{The panels (a) and (b) show two ways of linking EZMs in a $\zt$ or $\zth$ chain with $n_T = 2$ to form localized configurations.}
    \label{NT2_cpl} 
\end{figure}

\begin{figure}[!t]
    \centering 
    \centerline{\includegraphics[width=0.95\linewidth]{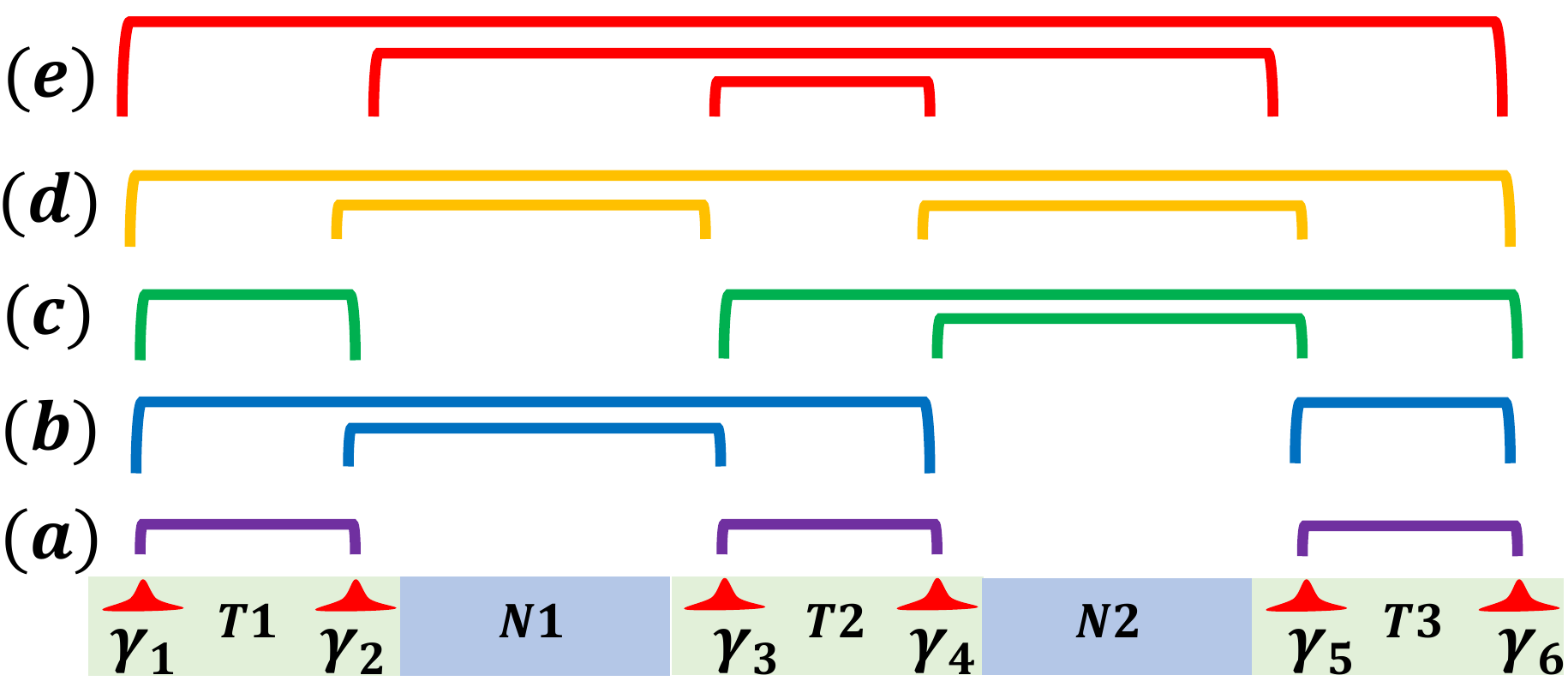}}
    \caption{The panels (a)-(e) show various ways of linking EZMs in a $\zt$ or $\zth$ chain with $n_T = 3$ to form localized configurations.}
    \label{NT3_cpl} 
\end{figure}

\indent In cases (a) and (b), the EZMs are localized at the interfaces, according to Fig.~\ref{NT2_cpl}(a) and (b), respectively. However, one might ask what happens if the $\D$'s are comparable. Let us assume, for the moment, that $\D_{T1} \simeq \D_{T2} \simeq \D_{N1}$. Unfortunately, in such cases, the DHE method does not work well, as its main assumption is that some couplings are significantly smaller or larger than the rest (or at least than the neighboring couplings at each step).
As shown in Ref.~\cite{narozhny2017majorana}, when the $\D$ values are comparable, the zero modes (ZMs) become "delocalized", meaning they are spread across multiple sites. For instance, one ZM may appear as a weighted combination of $\gamma_1$ and $\gamma_3$. However, the primary goal of topological quantum computation is to leverage the localized nature of EZMs, which ensures robustness against errors. Delocalization, on the other hand, compromises this robustness. Therefore, in this study, we focus solely on cases with localized ZMs.
We observed that the number of localized configurations, denoted by $n_C$, for $n_T = 2$ is two. Our investigations further revealed that, in general, this pattern follows the Catalan numbers
\begin{eqnarray}
\label{n_c_N_T}
n_C(n_T) = \frac{(2n_T)!}{(n_T + 1)! n_T!}.
\end{eqnarray}
The central insight leading to this result is that the lines connecting different pairs of EZMs must not intersect. For instance, with $n_T = 3$, we identified five possible localized configurations, as illustrated in Fig.~\ref{NT3_cpl}.

\begin{figure}[t]
    \centering 
    \centerline{\includegraphics[width=0.79\linewidth]{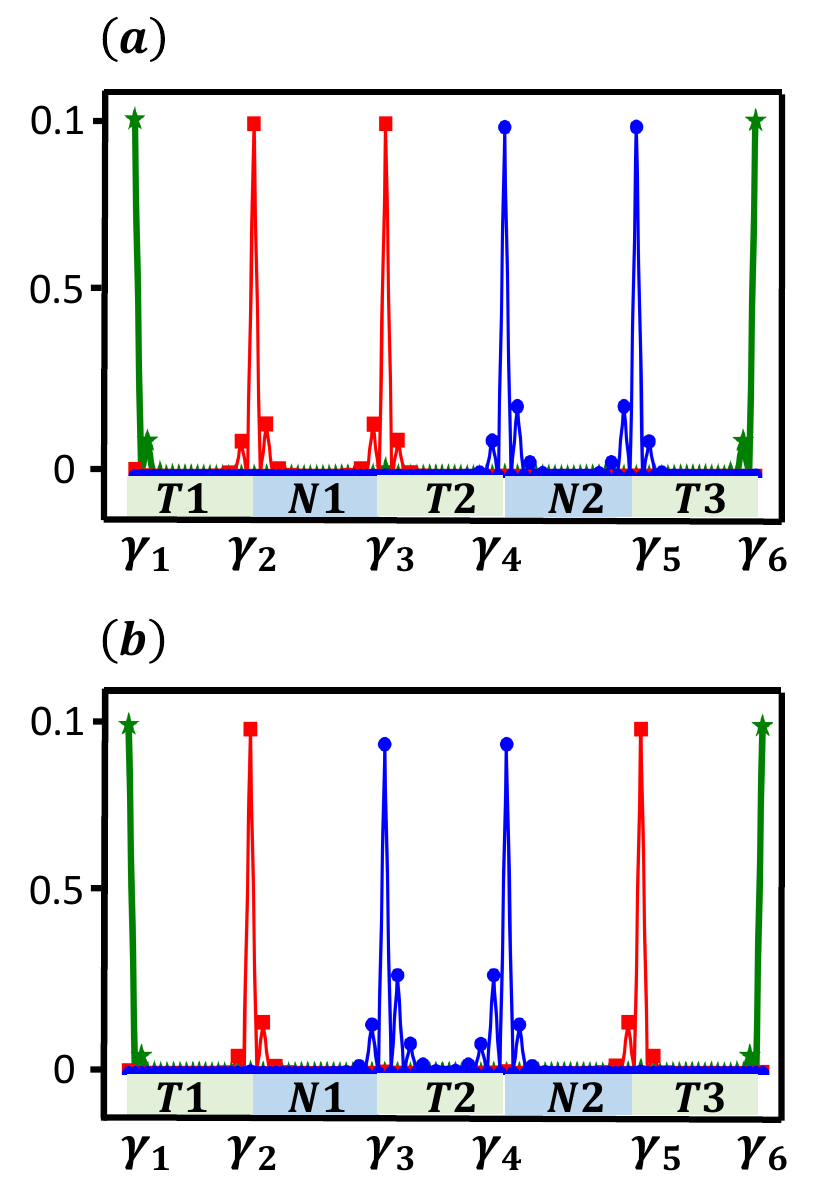}}
    \caption{In panels (a) and (b), the green stars, red squares, and blue circles represent the single-body wave functions of the first, second, and third excited states, respectively. Panel (a) corresponds to configuration (d), and panel (b) to configuration (e) in the $\zt$ chain with $N_T = 3$ discussed in the main text.}
    \label{NT3_DHE} 
\end{figure}

\indent Now, let us investigate our method for $n_T = 3$ and test it against numerical results. We will go over two examples that illustrate the method for the rest. Let us begin with the first set of parameters leading to the configuration shown in Fig.~\ref{NT3_cpl}(d): $L_{T_i} = L_{N_i} = 10 ; \forall i$, $J_{T_i} = J_{N_i} = 1 ; \forall i$, $f_{T1} = f_{T2} = f_{T3} = 0.1$, and $f_{N1} = 6.75$, $f_{N2} = 5$. Trivially, in this case,
\begin{center}
    (d) ~ $\D_{T1} = \D_{T2} = \D_{T3} = 1\times10^{-10} \ll \D_{N1} = 5.09\times 10^{-9} \ll \D_{N2} = 1.02\times 10^{-7}$.
\end{center}
The approach involves utilizing the DHE method once again as follows. Given that $\D_{N2}$ significantly surpasses the other couplings, we infer that $\gamma_4$ and $\gamma_5$ primarily engage with a strength of $\e^{(d)}_{45} = \D_{N2} = 1.02 \times 10^{-7}$.
Continuing the process by decimating this pair, we find that $\gamma_3$ and $\gamma_6$ interact with an effective strength of $\D_{\text{eff}} = \frac{\D_{T2} \D_{T3}}{\D_{N2}} = 0.98 \times 10^{-13}$.
Upon comparing the remaining couplings ($\D_{\text{eff}} \ll \D_{T1} \ll \D_{N1}$) and identifying the maximum, we observe that $\gamma_2$ and $\gamma_3$ interact with a strength of $\e^{(d)}_{23} = \D_{N1} = 5.09 \times 10^{-9}$.
Subsequently, after decimating this pair, we are left with $\gamma_1$ and $\gamma_6$, which are compelled to interact with a strength of $\e^{(d)}_{16} = \frac{\D{T1} \D_{\text{eff}}}{\e^{(d)}_{23}} = \frac{\D_{T1} \D_{T2} \D_{T3}}{\D_{N1} \D_{N2}} = 1.91 \times 10^{-15}$.
Thus, we obtain
\begin{eqnarray}
\label{H_Z2_NT3_d}
H^{(d)}_{\zt,n_T=3} =
i \e^{(d)}_{16} \gamma_{1}\gamma_{6} +
i \e^{(d)}_{23} \gamma_{2}\gamma_{3} +
i \e^{(d)}_{45} \gamma_{4}\gamma_{5},
\end{eqnarray}
leading to the association of the state $\ket{\psi^{(d)}} = \ket{n_{16},n_{23},n_{45}}$.
The numerical results obtained using the Exact Diagonalization method (see Appendix.~\ref{Apx_Exact} for details) show that $\e^{(d)}_{16} = 2.13 \times 10^{-15} \ll \e^{(d)}_{23} = 4.93 \times 10^{-9} \ll \e^{(d)}_{45} = 0.97 \times 10^{-7}$, in excellent agreement with our findings.
In Fig.~\ref{NT3_DHE}(a), we further present the numerical results for the three lowest single-body wave functions, each exhibiting the anticipated behavior.

\indent The second set of parameters, which leads to the configuration shown in Fig.~\ref{NT3_cpl}(e), was obtained by keeping all previous parameters unchanged while setting $f_{T1} = f_{T3} = 0.05$, $f_{T2} = 0.3$, and $f_{N1} = f_{N2} = 6.75$. We find that
\begin{center}
    (e) ~$\D_{T1} = \D_{T3} = 9.77\times10^{-14} \ll \D_{N1} = \D_{N2} = 5.09\times 10^{-9} \ll \D_{T2} = 5.90\times 10^{-6}$.
\end{center} 
By applying the DHE technique as before, it is evident that $\gamma_3$ and $\gamma_4$ primarily interact with a magnitude of $\epsilon^{(e)}_{34} = \Delta_{T2} = 5.90 \times 10^{-6}$.
Upon decimating this pair, we observe $\gamma_2$ and $\gamma_5$ interacting with $\D_{\eff} = \frac{\D_{N1} \D_{N2}}{\D_{T2}} = 4.39\times 10^{-12}$.
Given that $\D_{T1},~\D_{T3} \ll \D_{\eff}$, it follows that $\gamma_2$ and $\gamma_5$ should pair with strength $\e^{(e)}_{25} = \D_{\eff} = 4.39\times 10^{-12}$. 
Ultimately, $\gamma_{1}$ and $\gamma_{6}$ interact with $\e^{(e)}_{16} = \frac{\D_{T1} \D_{T3}}{\e^{(e)}_{25}} = \frac{\D_{T1} \D_{T2} \D_{T3}}{\D_{N1}\D_{N2}} = 2.17\times 10^{-15}$.
This confirms
\begin{eqnarray}
\label{H_Z2_NT3_e}
H^{(e)}_{\zt,n_T=3} =
i \e^{(e)}_{16} \gamma_{1}\gamma_{6} +
i \e^{(e)}_{25} \gamma_{2}\gamma_{5} +
i \e^{(e)}_{34} \gamma_{3}\gamma_{4},
\end{eqnarray}
resulting in the state $\ket{\psi^{(e)}} = \ket{n_{16},n_{25},n_{34}}$. 
The numerical results obtained using the Exact Diagonalization method indicate that $\e^{(e)}_{16} = 2.24\times 10^{-15} \ll \e^{(e)}_{25} = 4.29\times 10^{-12} \ll \e^{(e)}_{34} = 5.27\times 10^{-6}$ which aligns very well with our discoveries.
Figure~\ref{NT3_DHE}(b) further shows the numerical results for the three lowest single-body wave functions, each displaying the expected characteristics.

\indent Before delving into the $\zth$ model, we wish to underscore a few key points that apply to broader configurations. First, for cases where $n_T > 3$, the process aligns closely with the approach described for smaller $n_T$ values: we systematically identify the largest $\Delta$ and proceed with the steps outlined in the DHE method. This iterative procedure scales effectively with larger configurations, allowing consistent analysis across increasing complexity in the chains. Second, it is important to recognize that different parameter sets may lead to identical configurations. For example, in configuration (d), one could instead select $f_{N1} = 5$ and $f_{N2} = 6.75$ to yield the same structural pattern, highlighting a flexibility that can simplify parameter selection in certain cases.

\indent Finally, a critical question remains as to whether these different localized configurations can impact topological quantum computation. This question demands rigorous and comprehensive exploration, as identifying configurations that optimize qubit stability is central to the practical application of such systems. Addressing this issue will be a primary focus of our future work, aiming to provide deeper insight into the configurations that most effectively harness EZM robustness. Our objective in the present work has been to lay a strong foundation for these future investigations by developing a systematic approach for analyzing the low-energy physics of nonuniform chains. To the best of our knowledge, no previous work has provided a derivation with this level of specificity and applicability, which we believe will be essential for advancing the understanding of finite-size effects and the configuration landscape in parafermion systems. 

\indent These principles apply to the $\zth$ case as well, illustrating the versatility of our approach.

\section{$\zth$ Symmetry}
\indent In this section, we aim to extend our findings to the $\zth$ model, with certain adaptations. We begin by considering the Hamiltonian of the $\zth$ chain, given by 
\begin{eqnarray}
H_{\zth} = \sum_{i = 1}^{n_T} H_{Ti} + \sum_{i = 1}^{n_N} H_{Ni},
\end{eqnarray}
where now
\begin{eqnarray}
\label{H_Z3_TR}
H_{Ti} = 
&-& J_{Ti}\sum^{\vek{r}_{Ti} + L_{Ti}-1}_{j =  \vek{r}_{Ti} +1} {( \alpha_{Ti} \bar{\omega}\eta^{b \dagger}_{j} \eta^{a}_{j+1} 
+ h.c.)}\nonumber\\
&-& f_{Ti} \sum^{\vek{r}_{Ti} + L_{Ti}}_{j = \vek{r}_{Ti}+1} {( \hat{\alpha}_{Ti} \bar{\omega}\eta^{a \dagger}_{j} \eta^{b}_{j}
+ h.c. )} ,\nonumber\\
H_{Ni} = 
&-& J_{Ni}\sum^{\vek{r}_{Ni} + L_{Ni}}_{j = \vek{r}_{Ni}} {( \alpha_{Ni} \bar{\omega}\eta^{b \dagger}_{j} \eta^{a}_{j+1} + h.c. )} \nonumber\\
&-& f_{Ni} \sum^{\vek{r}_{Ni} + L_{Ni}}_{j = \vek{r}_{Ni}+1} {( \hat{\alpha}_{Ni} \bar{\omega}\eta^{a\dagger}_{j} \eta^{b}_{j} + h.c. )}.
\end{eqnarray}
In this case, the couplings $J$ and $f$ are real and non-negative, and $\alpha = e^{-i\phi}$ and $\hat{\alpha} = e^{-i\hat{\phi}}$ (with $\phi, \hat{\phi} \in [0, \pi/3)$) represent the chirality factors \cite{fendley2012parafermionic, fendley2014free, jermyn2014stability, alicea2016topological, zhuang2015phase}. The operators $\eta_i$ are generalizations of Majorana fermions with the properties $\eta^3 = I$, $\eta^{\dagger} = \eta^2$, and $\eta_i \eta_{j>i} = \omega \eta_j \eta_i$, where $\omega = e^{2\pi i / 3}$. The parameters $n_T$, $n_N = n_T - 1$, $L_{Ti}$, $L_{Ni}$, and $L$ retain their previous meanings.

\indent The DHE method is applicable to this model, as demonstrated for a single-region chain in the T phase in Ref.~\cite{benhemou2023universality}. In the Appendix.~\ref{Apx_DHE}, we show its applicability to the N phase. By following procedures analogous to those used for the $\zt$ symmetry, we obtain
\begin{eqnarray}
\label{H_Z3_eff}
H^{\eff}_{\zth} \approx
&-& \sum^{n_T}_{i=1} \D_{Ti} (\beta_{Ti}\bar{\omega} \eta^{\dagger}_{2i-1} \eta_{2i} + h.c.) \nonumber \\
&-& \sum^{n_N}_{i=1} \D_{Ni} (\beta_{Ni}\bar{\omega}\eta^{\dagger}_{2i} \eta_{2i+1} + h.c. ),
\end{eqnarray}
where $\eta$'s refer to the EZMs residing at the interfaces, and
\begin{eqnarray}
\label{D_Z3}
\D_{Ti} &=& f_{Ti}\Big(\frac{2\cos\phi_{Ti}}{(2\cos\phi_{Ti})^2-1} \times \frac{f_{Ti}}{J_{Ti}}\Big)^{L_{Ti-1}},\nonumber\\
\D_{Ni} &=& J_{Ni}\Big(\frac{2\cos\hat{\phi}_{Ni}}{(2\cos\hat{\phi}_{Ni})^2-1} \times \frac{J_{Ni}}{f_{Ni}}\Big)^{L_{Ni}}.
\end{eqnarray}
It is important to note that for certain angles, the $\D$'s do not diverge during the process. The effective chirality factors are now given by $\beta_{Ti} = e^{-i\theta_{Ti}}$ and $\beta_{Ni} = e^{-i\theta_{Ni}}$, where
\begin{eqnarray}
\label{Angle_Z3}
\theta_{Ti} &=& L_{Ti}\hat{\phi}_{Ti} ~~~~~~~~~~~~\mod \pi/3,\nonumber\\
\theta_{Ni} &=& (L_{Ni}+1)\phi_{Ni} ~\mod \pi/3.
\end{eqnarray}

\begin{figure}
    \centering 
    \centerline{\includegraphics[width=0.77\linewidth]{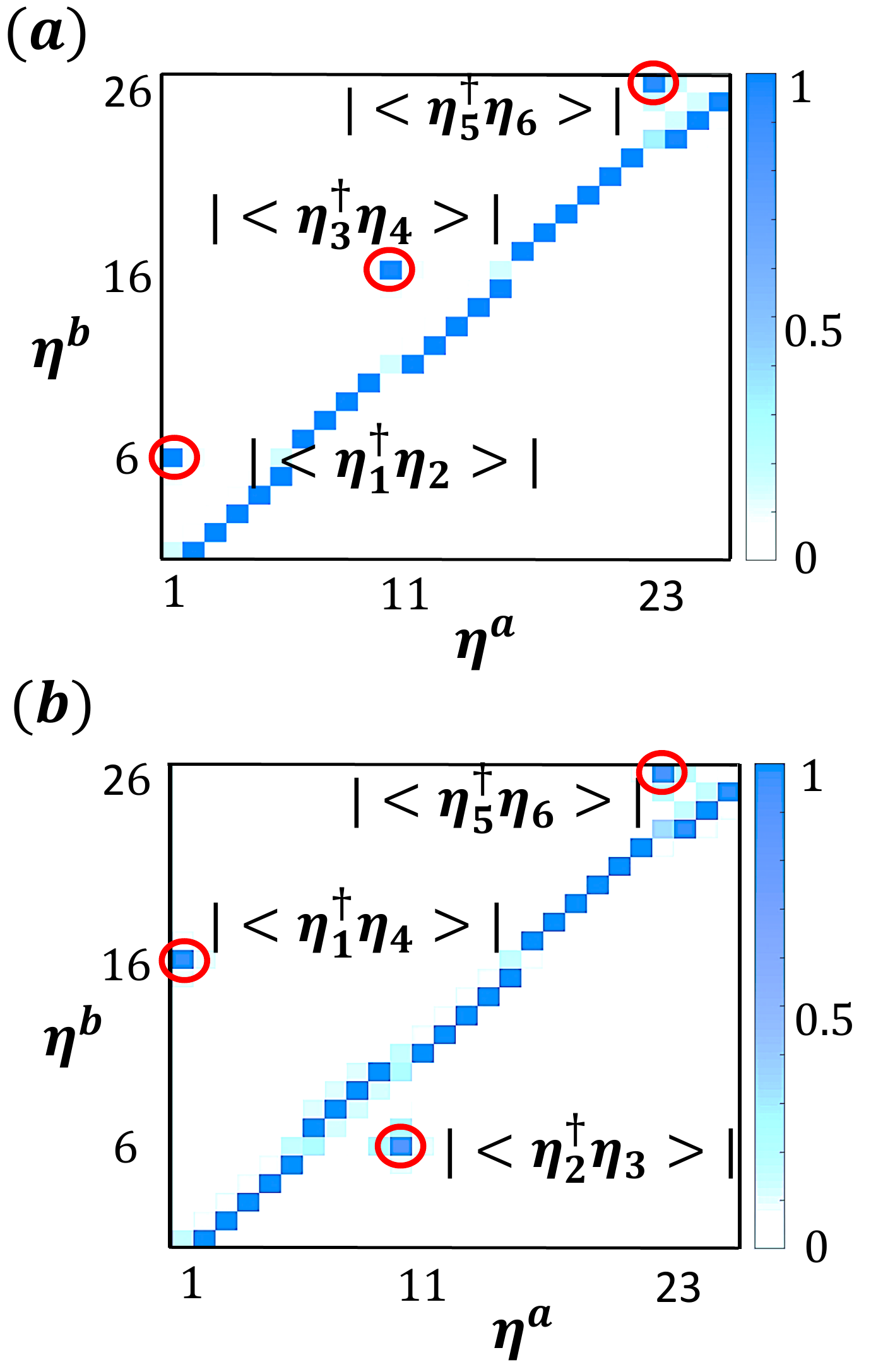}}
    \caption{Heatmap illustrating correlation functions in a $\zth$ parafermion chain with $n_T=3$. Red circles highlight long-range correlations among coupled EZMs. Indices on the axes correspond to real chain indices, indicating the locations of anticipated EZMs.}
    \label{Gr} 
\end{figure}
\begin{figure}
    \centering 
    \centerline{\includegraphics[width=0.8\linewidth]{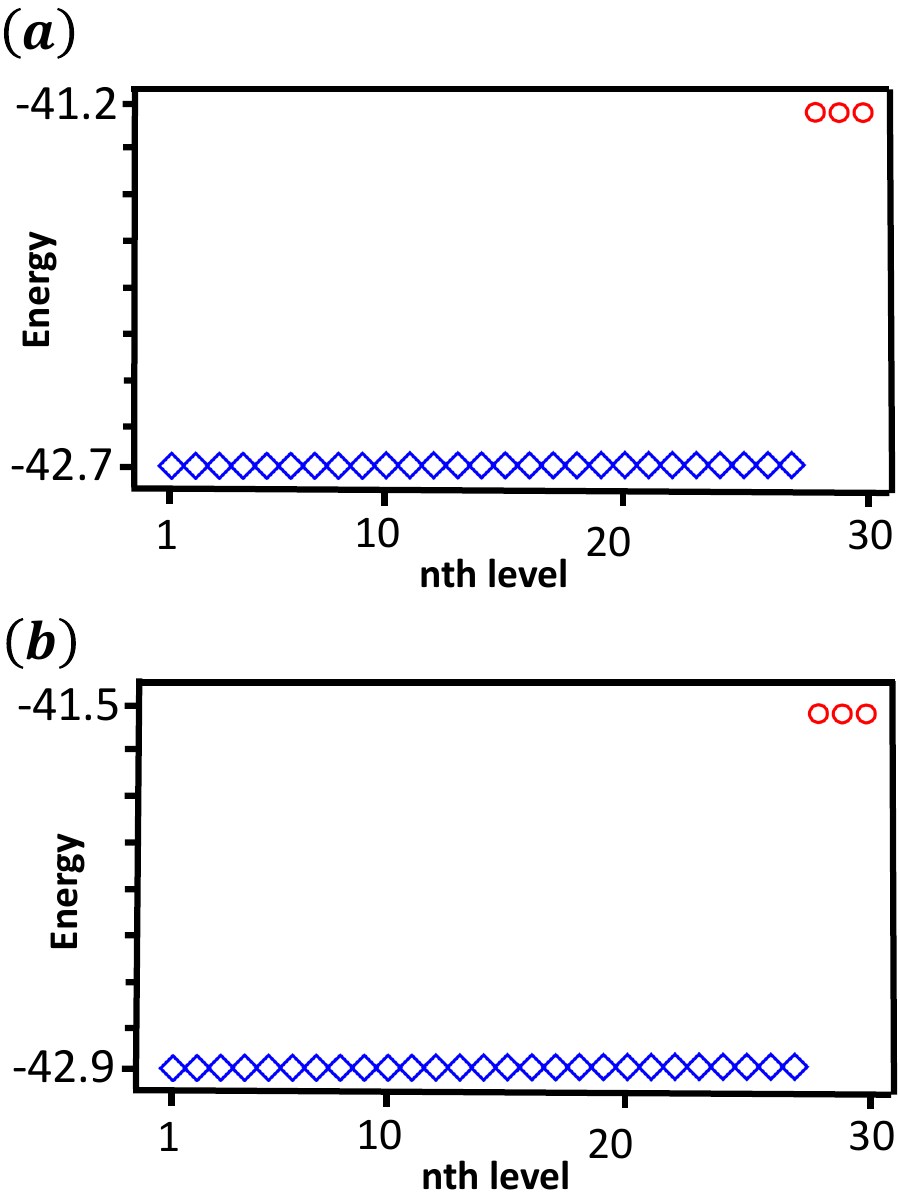}}
    \caption{(a) and (b) exhibit a clear gap between the many-body ground state manifold and the first excited states associated with parameter sets (a) and (b) in the $\zth$ examples, respectively.}
    \label{Gap_Z3} 
\end{figure}

\indent When $n_T = 1$ and, for example, $\phi = \hat{\phi} = 0$, we find that $H^{\eff}_{\zth} \approx \D_{T1} (\bar{\omega} \eta^{\dagger}_{1} \eta_{2} + h.c.)$, with $\D_{T1} = f_{T1} \left( \frac{2 f_{T1}}{3J_{T1}} \right)^{(L_{T1}-1)}$. This result is in agreement with Ref.\cite{jermyn2014stability} (particularly Eq. 15, although note that due to different model definitions, the factor of $2$ is absent there). It is also worth noting that, at $\phi = \hat{\phi} = 0$, the EZMs display weak characteristics, whereas at $\phi = \hat{\phi} = \pi/6$, the model exhibits super-integrability, leading to strong EZMs\cite{fendley2012parafermionic, fendley2014free, jermyn2014stability, alicea2016topological}.

\indent Our investigations for $n_T > 1$ revealed results similar to those for the $\zt$ case. Specifically, we identified some localized configurations, the number of which follows Eq.(\ref{n_c_N_T}), as well as some delocalized cases. However, our focus here remains on the localized configurations, as before. We will review two examples for $n_T = 3$, which give rise to the configurations shown in Fig.\ref{NT3_cpl} (a) and (b). We do not examine $n_T = 2$ separately, as we expect the patterns to resemble those in Fig.~\ref{NT2_cpl} by setting $L_{N2} = L_{T3} = 0$, effectively eliminating EZMs numbered $5$ and $6$. Thus, these examples are representative for both $n_T = 2$ and $n_T = 3$.

\indent Here, we consider two sets of parameters yielding configurations similar to those shown in Fig.~\ref{NT3_cpl} (a) and (b), respectively. In the first set, we have $L_{T1} = L_{T2} = L_{N2} = 6$, $L_{N1} = L_{T3} = 4$, $J_{Ti} = f_{Ni} = 1 ; \forall i$, $f_{T1} = f_{T2} = J_{N2} = 0.1$, $f_{T3} = 0.2$, and $J_{N1} = 0.05$. We also assume $\phi_{Ti} = \hat{\phi}_{Ti} = \pi/6 ; \forall i$ and $\phi_{Ni} = \hat{\phi}_{Ni} = 0 ; \forall i$. In this setup, we have
\begin{center}
    (a) ~$\D_{N2} = 8.78\times 10^{-9} \ll \D_{N1} = 6.17\times 10^{-8} \ll \D_{T1} = \D_{T2} = 4.87 \times 10^{-7} \ll \D_{T3} = 1.04\times 10^{-3}$.
\end{center}
\indent Applying the DHE method and following the procedure used for $\zt$, based on the competition among the $\D$'s, it is evident that $\eta_5$ and $\eta_6$ pair first, followed by $\eta_3$ and $\eta_4$, and finally $\eta_1$ and $\eta_2$. Thus,
\begin{eqnarray}
\label{H_Z3_NT3_a}
H^{(a)}_{\zth,n_T = 3} =
&-&\e^{(a)}_{12} (\bar{\omega}\eta^{\dagger}_{1}\eta_{2} + h.c.)
-\e^{(a)}_{34} (\bar{\omega}\eta^{\dagger}_{3}\eta_{4} + h.c.) \nonumber\\
&-&\e^{(a)}_{56} (\bar{\omega}\eta^{\dagger}_{5}\eta_{6} + h.c.),
\end{eqnarray}
where $\e^{(a)}_{12} = \D_{T1},~\e^{(a)}_{34} = \D_{T2}$, and $\e^{(a)}_{56} = \D_{T3}$.

\indent For the second set, we maintained the same parameters as in the example above but set $J_{N1} = 0.3$ and applied the DHE method. We obtain $\D_{N2} = 8.78 \times 10^{-9}$. Here, we observe that
\begin{center}
    (b) ~$\D_{N2} \ll \D_{T1} = \D_{T2} = 4.87 \times 10^{-7} \ll \D_{N1} = 4.80\times 10^{-4} < \D_{T3} = 1.04\times 10^{-3}$
\end{center}
and thus
\begin{eqnarray}
\label{H_Z3_NT3_b}
H^{(b)}_{\zth,N_T = 3} =
&-&\e^{(b)}_{14} (\bar{\omega}\eta^{\dagger}_{1}\eta_{4} + h.c.)
-\e^{(b)}_{23} (\bar{\omega}\eta^{\dagger}_{2}\eta_{3} + h.c.) \nonumber\\
&-&\e^{(b)}_{56} (\bar{\omega}\eta^{\dagger}_{5}\eta_{6} + h.c.),
\end{eqnarray}
where $\e^{(b)}_{14} = \frac{2}{3} \frac{\D_{T1}\D_{T2}}{\D_{N1}} = 3.30\times 10^{-10}$, $\e^{(b)}_{23} = \D_{N1}$, and $\e^{(b)}_{56} = \D_{T3}$.

\indent Unfortunately, standard methods for diagonalizing $H_{\zt}$ cannot be applied to $H_{\zth}$ \cite{fendley2012parafermionic, cobanera2014fock}. Thus, alternative approaches are needed to numerically identify coupled EZMs. In Appendix \ref{Apx_Gr}, we demonstrate that the Green's function $G_{k,n} = |\braket{\eta^{a\dagger}_{k} \eta^{b}_n}|$ effectively serves this purpose. Specifically, long-range correlations reveal which EZMs are coupled. Numerical results for the examples in Fig.\ref{Gr}, obtained using DMRG, align well with our expectations. Furthermore, the plots in Fig.\ref{Gap_Z3} confirm nearly 27-fold degenerate ground states with a finite gap to excited states in the aforementioned examples. These findings suggest that the conclusions drawn from $\zt$ symmetry can be extended to the $\zth$ case.

\begin{figure*}
    \centering 
    \centerline{\includegraphics[width=0.7\linewidth]{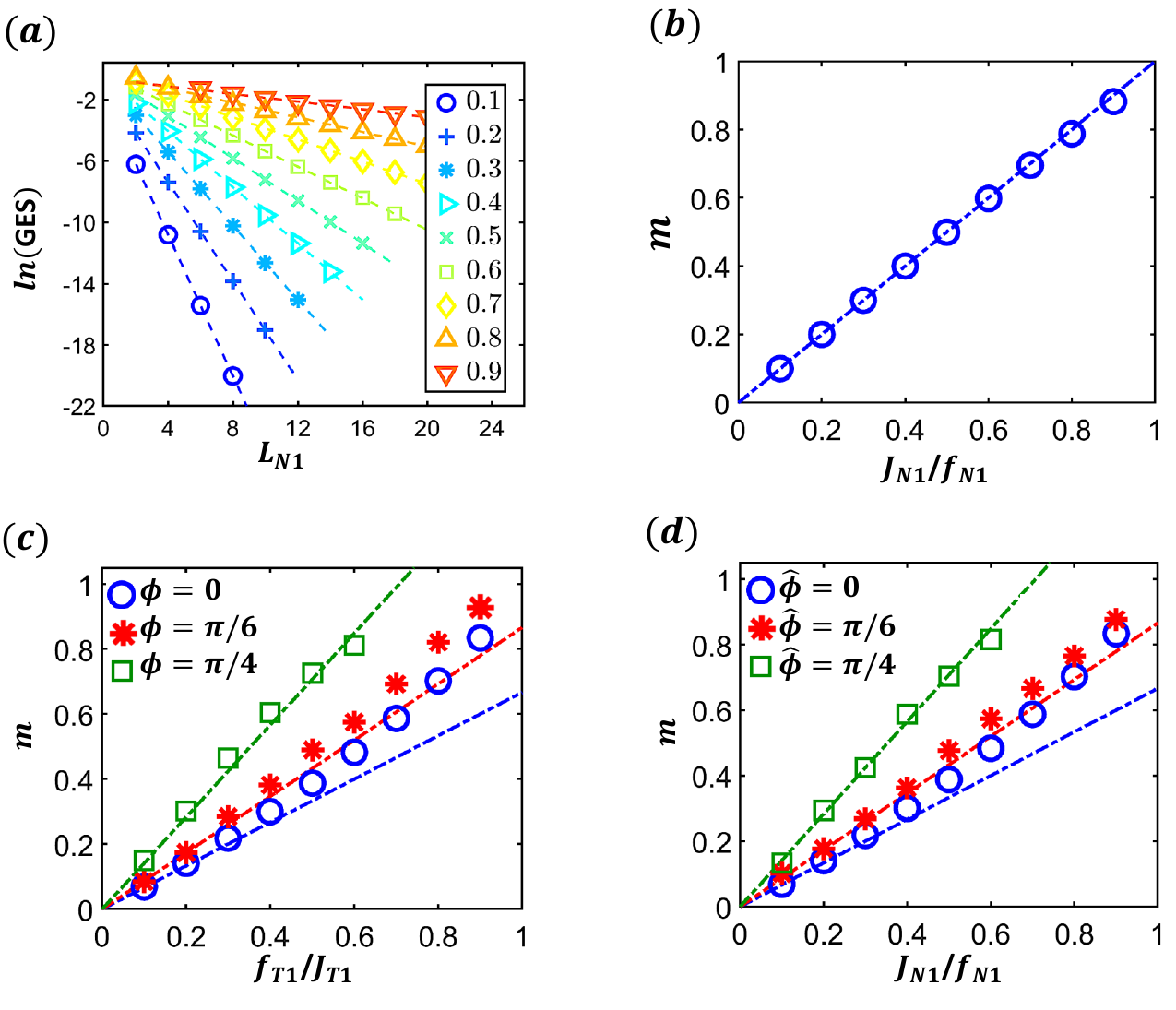}}
    \caption{In a $\zt$ chain with $n_T = 2$, (a) shows that varying $L_{N1}$ while keeping other parameters constant results in $\text{GES} \propto (m)^{L_{N1}}$, akin to $\D_{N1}$, while (b) demonstrates $m = J_{N1}/f_{N1}$. Similar analyses were conducted for a $\zth$ chain, confirming the validity of $\text{GES} \propto (m)^{L}$. The effect of chirality on $m$ becomes evident for small $J$ and $f$ ratios, as shown in (c) for the T phase and (d) for the N phase, following Eq. (\ref{D_Z3}). As ratios increase, the chirality effect diminishes, and $m$ primarily depends on $J/f$.}
    \label{Dlt} 
\end{figure*}
\section{ground state energy splitting analysis}

\indent Here, we show that the $\D$ values reflect the ground state energy splitting (GES) due to finite-size effects in each region, leading to the definition of critical lengths for them.
In $\zt$ chains, when $n_T = 1$, we confirm the Kitaev result \cite{kitaev2001unpaired} from Eq. (\ref{H_Z2_eff}), affirming the link between $\D_{T1}$ and GES. It's crucial to ensure that $L_{T1} \gg \xi_{T1}$ to maintain well-separated EZMs, defining the critical length as $\xi_{T1} = 1/|\ln(f_{T1}/J_{T1})|$.
For $n_T = 2$, our numerical results, depicted in Fig. \ref{Dlt}(a, b), show that GES $\propto (\frac{J_{N1}}{f_{N1}})^{L_{N1}}$, while ensuring $L_{Tn} \gg \xi_{Tn}$ for $n=1,2$, with all other parameters fixed. This procedure can generally be repeated for other regions, yielding similar outcomes. Consequently, critical lengths can be defined for each region as:
\begin{eqnarray} 
\label{CL} 
\xi = 1/|\ln(q)|, 
\end{eqnarray}
where $q = \frac{f}{J}$ ($q = \frac{J}{f}$) for T (N) regions. Thus, to guarantee well-separated EZMs, it's necessary to have $L_{Ti} \gg \xi_{Ti}$ and $L_{Ni} \gg \xi_{Ni}$ for all $i$.

\indent In $\zth$ chains, the GES analysis depicted in Fig.~\ref{Dlt}(c, d) mirrors the findings in the $\zt$ case, exhibiting an anticipated trend for small ratios of $J$ and $f$, albeit with deviations for larger ones. Once more, ensuring well-separated EZMs necessitates $L_{Ti} \gg \xi_{Ti}$ and $L_{Ni} \gg \xi_{Ni}$ for all $i$, utilizing Eq. (\ref{CL}) with $q = \frac{2\cos\phi}{(2\cos\phi)^2-1} \times \frac{f}{J}$ ($q = \frac{2\cos\hat{\phi}}{(2\cos\hat{\phi})^2-1} \times \frac{J}{f}$) for T (N) regions.

\indent As easily verified, the condition $L \gg \xi$ is met for all regions in the examples studied throughout this paper

\section{Conclusion}

\indent In this study, we introduced a novel approach based on the decimation of the highest energy term to derive the low-energy Hamiltonian for $\zt$ and $\zth$ symmetric parafermion chains, encompassing a range of topological and normal regions. 
By applying the DHE method, we confirmed previously established results for the $n_T = 1, 2$ $\zt$ chains and $n_T = 1$ $\zth$ chains. We extended our analysis to the $n_T = 3$ case for both $\zt$ and $\zth$ symmetries, validating the analytical results with numerical computations.
Furthermore, our investigation of the ground state energy splitting provided a crucial metric for ensuring well-separated EZMs in finite-size systems, which was vital for their stability.
Overall, the method we presented proved highly versatile and can be readily adapted to investigate higher symmetries, paving the way for further exploration in the field. Questions regarding how these findings might benefit topological quantum computation require further investigation and is the focus of our future work.

\begin{acknowledgments}
We are grateful to Abolhassan Vaezi and Reza Asgari for their valuable and insightful discussions.\\
\end{acknowledgments}

\indent \textit{\textbf{Author contributions.}}\,---M.-S.V. conceived the project, devised the methodology, conducted the investigation, analyzed the data, generated the visualizations, and drafted the original manuscript. M.M.N. contributed to the investigation, visualizations, and assisted in revising the draft. M.-S.V. supervised the project.

\appendix

\section{Exact Diagonalization Method}
\label{Apx_Exact}
\indent The Eq.(\ref{H_Z2_main}) can be rewritten as
\begin{eqnarray}
H_{\zt} = \vek{\gamma}^{\dagger} A \vek{\gamma},
\end{eqnarray}
where $A$ is a Hermitian matrix and $\vek{\gamma}_{2m-1} = \gamma^{a}_m,~\vek{\gamma}_{2m} = \gamma^{b}_m$. 
We can diagonalize it using a unitary transformation $U$ as
\begin{eqnarray}
\label{H_Z2_diag}
H^{\text{diag}}_{\zt} = \sum^{L}_{k= -L, k\neq 0} \e_{k} n_{k},
\end{eqnarray}
where $\e = U^{\dagger}AU$ and $n_{k} = c^{\dagger}_{k} c_{k}$ represent the single-particle excitation energy and operator, respectively. The quasi-fermionic annihilation operators are given by $c = U^{\dagger} \vek{\gamma}$. One should note that $\e_{-k} = -\e_{k}$.
Hence, the eigenvalues of matrix $A$ yield the desired energies, while the columns of matrix $U$ furnish the corresponding eigenvectors. These eigenvectors consist of both real and imaginary elements that convey information about the edge zero modes residing at the left and right sides of each T region for the lowest-lying modes, respectively.

\section{DHE Method}
\label{Apx_DHE}
\begin{figure}[b]
    \centering 
    \centerline{\includegraphics[width=0.6\linewidth]{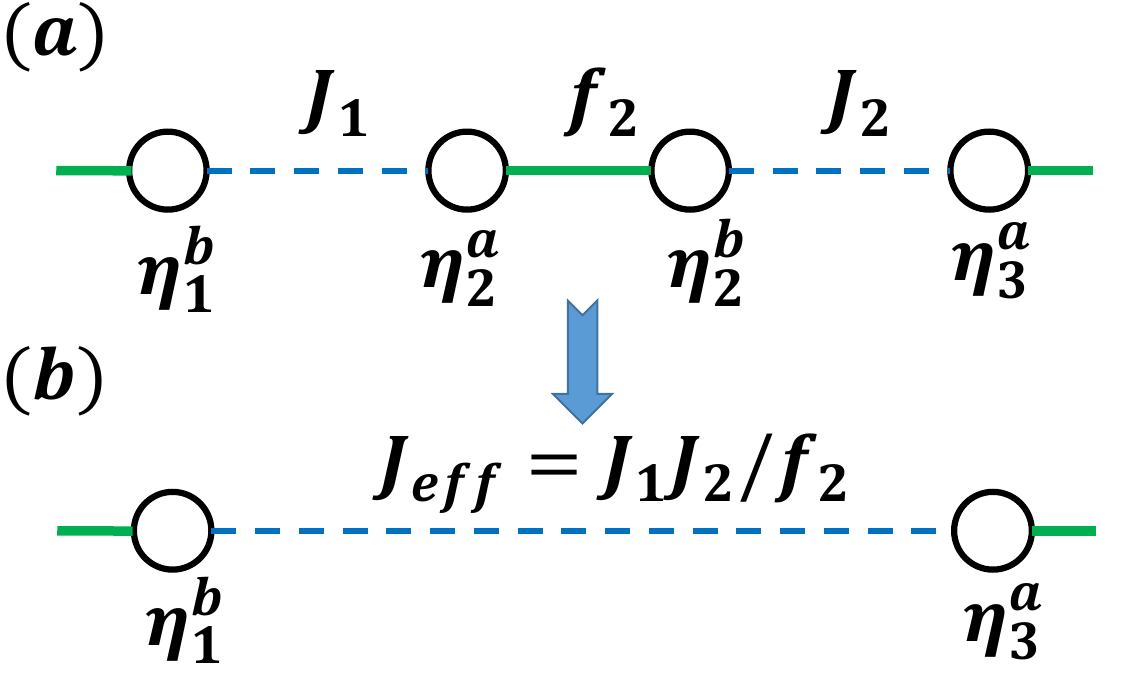}}
    \caption{Schematic illustration of the DHE method. Decimating the $J$ couplings $\gamma^{b}_{1}$ and $\gamma^{a}_{2}$ in panel (a) associated with a $\zt$ symmetric chain leads to the configuration illustrated in panel (b). Decimating the $f_2$ coupling between $\eta^{a}_{2}$ and $\eta^{b}_{2}$ in panel (c) associated with a $\zth$ symmetric chain leads to the depiction shown in panel (d).}
    \label{DHE_Z3} 
\end{figure}

We stated that the DHE method is an iterative process that involves eliminating the most energetic term within the Hamiltonian and replacing it with effective longer-range interactions using second-order perturbation theory at each iteration.
Here, we show a part of the procedure of applying this method to a $\zth$ single-region chain in the N phase.
We compute the effective term depicted in Fig.~\ref{DHE_Z3} when $f_2$ dominates over the other couplings.
The Hamiltonian of this part is given by $H = H_{f}+H_{J}$, where
\begin{eqnarray}
\label{H_Jf}
H_f &=& -f_2 \bar{\omega} \hat{\alpha} \eta^{a\dagger}_{2}\eta^{b}_{2} + h.c. \nonumber\\
H_J &=& -J_1 \bar{\omega} \alpha \eta^{b\dagger}_{1}\eta^{a}_{2}
- J_2 \bar{\omega} \alpha \eta^{b\dagger}_{2}\eta^{a}_{3} + h.c.,
\end{eqnarray}
with $\omega = e^{i\frac{2\pi}{3}}$, $\bar{\omega} = \omega^{2}$, $\alpha = e^{-i\phi}$, $\hat{\alpha} = e^{-i\hat{\phi}}$, and $\phi,\hat{\phi} \in [0,\pi/3)$. Utilizing the generalized Jordan-Wigner transformation, we obtain
\begin{eqnarray}
\label{JWT}
\eta^{b}_1 = \omega (\sigma \tau \otimes I \otimes I),
~\eta^{a}_2 = \tau \otimes \sigma \otimes I, \nonumber\\
\eta^{b}_2 = \omega (\tau \otimes \sigma \tau \otimes I),
~\eta^{a}_3 = \tau \otimes \tau \otimes \sigma, 
\end{eqnarray}
where
\begin{eqnarray}
\sigma = 
\left(
\begin{matrix}
    1 & 0 & 0 \\
    0 & \omega & 0 \\
    0 & 0 & \omega^2 \\
\end{matrix} 
\right),
\tau = 
\left(
\begin{matrix}
    0 & 0 & 1 \\
    1 & 0 & 0 \\
    0 & 1 & 0 \\
\end{matrix} 
\right),
I = 
\left(
\begin{matrix}
    1 & 0 & 0 \\
    0 & 1 & 0 \\
    0 & 0 & 1 \\
\end{matrix} 
\right).
\end{eqnarray}
From Eqs.(\ref{H_Jf}) and (\ref{JWT}):
\begin{eqnarray}
H_f = &-& f_2 \hat{\alpha} (I \otimes \tau \otimes I) + h.c., \nonumber\\
H_J = &-& J_1 \alpha (\sigma^{\dagger}\otimes \sigma \otimes I)
- J_2 \alpha (I\otimes \sigma^{\dagger}\otimes \sigma) + h.c.,\nonumber\\
\end{eqnarray}
If $\ket{0}$, $\ket{1}$, and $\ket{2}$ denote the eigenvectors of $\sigma$ with corresponding eigenvalues $0$, $\omega$, and $\omega^2$, respectively, then the eigenvectors of $\tau$'s will be:
\begin{eqnarray}
\ket{0}_{\tau} &=& \frac{1}{\sqrt{3}}(\ket{0} + \ket{1} + \ket{2}),\nonumber\\
\ket{1}_{\tau} &=& \frac{1}{\sqrt{3}}(\omega \ket{0} + \omega^2 \ket{1} + \ket{2}),\nonumber\\
\ket{2}_{\tau} &=& \frac{1}{\sqrt{3}},(\omega^2 \ket{0} + \omega \ket{1} + \ket{2})
\end{eqnarray}
associated with eigenvalues $0$, $\omega$, and $\omega^2$, respectively. It is easy to show that the ground state energy $E_{gs}$, first excited state $E_{e1}$, and second excited state $E_{e2}$ of $H_f$ are:
\begin{eqnarray}
E_{gs} &=& -2f_2 \cos\hat{\phi}, \nonumber\\
E_{e1} &=& 2f_2 \cos(\pi/3 + \hat{\phi}), \nonumber\\
E_{e2} &=& 2f_2 \cos(\pi/3 -\hat{\phi}),
\end{eqnarray}
The corresponding eigenvectors are each $9$-fold degenerate,
\begin{eqnarray}
\ket{gs}_{i,j} &=& \ket{i}\otimes \ket{0}_{\tau} \otimes \ket{j}\nonumber\\
\ket{e1}_{i,j} &=& \ket{i}\otimes \ket{1}_{\tau} \otimes \ket{j}, \nonumber \\
\ket{e2}_{i,j} &=& \ket{i}\otimes \ket{2}_{\tau} \otimes \ket{j},
\end{eqnarray}
where $i,j = 0,1,2$. Since the first order perturbation vanishes, we use the second-order perturbation theory to find the effective $H_J$ projected to the ground state of $H_f$:
\begin{eqnarray}
H^{\eff}_{ij,kl} = \sum_{m=,1,2}\sum_{i',j'}\frac{{}_{ij}\bra{gs}H_J\ket{em}_{i'j'}\bra{em}H_J\ket{gs}_{kl}}{E_{gs} - E_{em}}.
\end{eqnarray}
Using the fact that $\sum_{x = 0,1,2}\ket{x}\bra{x} = I$, it easy to show that:
\begin{eqnarray}
H^{\eff} &=& \sum_{m=1,2}\frac{B_{0,m}B_{m,0}}{E_{gs} - E_{em}}
\end{eqnarray}
where $B_{m,0} = B_{0,m}^{\dagger}$ and
\begin{eqnarray}
B_{0,m} = (I\otimes {}_{\tau}\bra{0} \otimes I)H_J (I\otimes \ket{m}_{\tau}\otimes I).
\end{eqnarray}
For $B_{0,m}$ we have:
\begin{eqnarray}
B_{0,m} &=& (I\otimes {}_{\tau}\bra{0} \otimes I)\Big[-J_1 (\alpha \sigma^{\dagger}\otimes \sigma \otimes I + \alpha^{*} \sigma\otimes \sigma^{\dagger} \otimes I) \nonumber\\
&-&J_2 ( \alpha I\otimes \sigma^{\dagger}\otimes \sigma + \alpha^{*} I\otimes \sigma\otimes \sigma^{\dagger}) \Big](I\otimes \ket{m}_{\tau}\otimes I) \nonumber\\
&=& -J_1 (\alpha \sigma^{\dagger}\otimes {}_{\tau}\bra{0} \sigma \ket{m}_{\tau} \otimes I + \alpha^{*} \sigma \otimes  {}_{\tau}\bra{0}\sigma^{\dagger}\ket{m}_{\tau} \otimes I) \nonumber\\
&-& J_2 ( \alpha I\otimes {}_{\tau}\bra{0} \sigma^{\dagger} \ket{m}_{\tau} \otimes \sigma + \alpha^{*} I\otimes {}_{\tau}\bra{0} \sigma \ket{m}_{\tau} \otimes \sigma^{\dagger})
\nonumber\\
\end{eqnarray}
Using the facts:
\begin{eqnarray}
\sigma^{\dagger} \ket{1}_{\tau} &=&  \omega \ket{0}_{\tau},~~~~~
\sigma \ket{1}_{\tau} =  \omega^{2} \ket{2}_{\tau}, \nonumber\\
\sigma^{\dagger} \ket{2}_{\tau} &=&  \omega \ket{1}_{\tau},~~~~~
 \sigma \ket{2}_{\tau} =  \omega^{2} \ket{0}_{\tau},
\end{eqnarray}
we will have:
\begin{eqnarray}
B_{0,1} &=& -J_1 \alpha^{*} \omega \sigma \otimes I  -J_2 \alpha \omega I \otimes \sigma \nonumber\\
B_{0,2} &=& -J_1 \alpha \omega^{2} \sigma^{\dagger} \otimes I  -J_2 \alpha^{*} \omega^{2} I \otimes \sigma^{\dagger}.
\end{eqnarray}
We note that $B_{0,2} = B_{0,1}^{\dagger}$.
Putting all these together:
\begin{eqnarray}
B_{0,1}B_{1,0} &=& B_{0,2}B_{2,0} = (J^{2}_{1}+J^{2}_{2}) I \otimes I \nonumber\\
&+& J_{1}J_{2} \Big[\alpha \sigma^{\dagger}\otimes \sigma + h.c. \Big].\nonumber\\
\end{eqnarray}
Using the previously introduced generalized Jordan-Wigner transformation, it is straightforward to demonstrate (in the projected space):
\begin{eqnarray}
\sigma^{\dagger} \otimes \sigma 
= \bar{\omega} \eta^{\dagger b}_{1} \eta^{a}_{3}.
\end{eqnarray}
Ultimately:
\begin{eqnarray}
H^{\eff} &=& \Big[-\frac{J^{2}_{1}+J^{2}_{2}}{2f_2} -\frac{J_1J_2}{2f_2} \times (\alpha^{2} \bar{\omega} \eta^{\dagger b}_{1} \eta^{a}_{3} + h.c.) \Big] \nonumber\\
&\times& \Big[\frac{1}{\cos\hat{\phi}+\cos(\pi/3 +\hat{\phi})} + \frac{1}{\cos\hat{\phi}+\cos(\pi/3 -\hat{\phi})}\Big] \nonumber \\
&=& \Big[-\frac{J^{2}_{1}+J^{2}_{2}}{f_2} -\frac{J_1J_2}{f_2} \times (\alpha^{2} \bar{\omega} \eta^{\dagger b}_{1} \eta^{a}_{3} + h.c.) \Big] \nonumber\\
&\times& \frac{2\cos\hat{\phi}}{(2\cos\hat{\phi})^2 -1}.
\end{eqnarray}
One can repeat this process and decimate the remaining terms until arriving at the expressions presented for $\zth$ symmetry.
\\
\section{EZMs Correlation}
\label{Apx_Gr}

\indent Here, we justify the utilization of the Green's function $G_{kn} = \langle \eta^{a\dagger}_{k} \eta^{b}_n \rangle$ to discern which EZMs couple. For instance, in a scenario with $n_T = 2$ and zero chirality, where $\eta_1$ and $\eta_2$ form the first pair, and $\eta_3$ and $\eta_4$ form the second pair, the effective Hamiltonian would be as follows
\begin{eqnarray}
H = -\D^{0}_{1,2} \bar{\omega} \eta^{\dagger}_{1}\eta_{2}
-\D^{0}_{3,4} \bar{\omega}\eta^{\dagger}_{3}\eta_{4} + h.c.,
\end{eqnarray}
Utilizing the generalized Jordan-Wigner transformation, we obtain
\begin{eqnarray}
\eta_1 &=& \sigma \otimes I,
~\eta_2 = \omega (\sigma \tau \otimes I), \nonumber\\
\eta_3 &=& \tau \otimes \sigma,
~\eta_4 = \omega (\tau \otimes \sigma \tau).
\end{eqnarray}
We can rewrite $H$ as
\begin{eqnarray}
H = -\D^{0}_{1,2} (\tau \otimes I)
-\D^{0}_{3,4} (I \otimes \tau) + h.c..
\end{eqnarray}
Using the basis of $\tau$, i.e., $\ket{0}_{\tau}, \ket{1}_{\tau}, \ket{2}_{\tau}$, with eigenvalues $1,\omega,\omega^2$, respectively, and assuming the total parity is $0$, the state of the system would be $\ket{\psi} = \ket{0}_{\tau}\otimes\ket{0}_{\tau} \equiv \ket{0,0}_{\tau}$. Noting the fact that $\sigma \ket{m}_{\tau} \propto \ket{(m+1) \mod{3}}_{\tau}$, we can easily demonstrate that $G_{1,2}$ and $G_{3,4}$ are nonzero:
\begin{eqnarray}
G_{1,2} = |\bra{\psi} \eta^{\dagger}_1\eta_2 \ket{\psi}| &=& |{}_{\tau}\bra{0,0} \tau\otimes I \ket{0,0}_{\tau}| = 1, \nonumber\\
G_{3,4} = |\bra{\psi} \eta^{\dagger}_3\eta_4 \ket{\psi}| &=& |{}_{\tau}\bra{0,0} I\otimes \tau \ket{0,0}_{\tau}| = 1.
\end{eqnarray}
If we were to compute $G_{2,3}$ and $G_{1,4}$ instead, they would be zero:
\begin{eqnarray}
G_{2,3} &=& |\bra{\psi} \eta^{\dagger}_2\eta_3 \ket{\psi}| \nonumber\\
&=& |{}_{\tau}\bra{0,0} \sigma^{\dagger}\otimes \sigma \ket{0,0}_{\tau}| = |{}_{\tau}\langle 0,0|2,1\rangle_{\tau}| = 0,\nonumber\\
G_{1,4} &=& |\bra{\psi} \eta^{\dagger}_1\eta_4 \ket{\psi}| = |{}_{\tau}\bra{0,0} \sigma^{\dagger}\tau\otimes \sigma \tau \ket{0,0}_{\tau}|\nonumber\\
&=& |{}_{\tau}\bra{0,0} (\sigma^{\dagger}\otimes \sigma)(\tau\otimes \tau) \ket{0,0}_{\tau}| = |{}_{\tau}\bra{0,0} \sigma^{\dagger}\otimes \sigma \ket{0,0}_{\tau}|\nonumber\\
&=& 0.\nonumber
\end{eqnarray}
Following the described procedure, it can be demonstrated that if $\eta_1$ and $\eta_4$ form the first pair in the system, and $\eta_2$ and $\eta_3$ form the second pair, then $G_{1,2} = G_{3,4} = 0$, while $G_{1,4}$ and $G_{2,3}$ are nonzero.

\nocite{apsrev41Control}
\bibliographystyle{apsrev4-1}

\begin{thebibliography}{39}%
\makeatletter
\providecommand \@ifxundefined [1]{%
 \@ifx{#1\undefined}
}%
\providecommand \@ifnum [1]{%
 \ifnum #1\expandafter \@firstoftwo
 \else \expandafter \@secondoftwo
 \fi
}%
\providecommand \@ifx [1]{%
 \ifx #1\expandafter \@firstoftwo
 \else \expandafter \@secondoftwo
 \fi
}%
\providecommand \natexlab [1]{#1}%
\providecommand \enquote  [1]{``#1''}%
\providecommand \bibnamefont  [1]{#1}%
\providecommand \bibfnamefont [1]{#1}%
\providecommand \citenamefont [1]{#1}%
\providecommand \href@noop [0]{\@secondoftwo}%
\providecommand \href [0]{\begingroup \@sanitize@url \@href}%
\providecommand \@href[1]{\@@startlink{#1}\@@href}%
\providecommand \@@href[1]{\endgroup#1\@@endlink}%
\providecommand \@sanitize@url [0]{\catcode `\\12\catcode `\$12\catcode `\&12\catcode `\#12\catcode `\^12\catcode `\_12\catcode `\%12\relax}%
\providecommand \@@startlink[1]{}%
\providecommand \@@endlink[0]{}%
\providecommand \url  [0]{\begingroup\@sanitize@url \@url }%
\providecommand \@url [1]{\endgroup\@href {#1}{\urlprefix }}%
\providecommand \urlprefix  [0]{URL }%
\providecommand \Eprint [0]{\href }%
\providecommand \doibase [0]{http://dx.doi.org/}%
\providecommand \selectlanguage [0]{\@gobble}%
\providecommand \bibinfo  [0]{\@secondoftwo}%
\providecommand \bibfield  [0]{\@secondoftwo}%
\providecommand \translation [1]{[#1]}%
\providecommand \BibitemOpen [0]{}%
\providecommand \bibitemStop [0]{}%
\providecommand \bibitemNoStop [0]{.\EOS\space}%
\providecommand \EOS [0]{\spacefactor3000\relax}%
\providecommand \BibitemShut  [1]{\csname bibitem#1\endcsname}%
\let\auto@bib@innerbib\@empty
\bibitem [{\citenamefont {Nielsen}\ and\ \citenamefont {Chuang}(2010)}]{nielsen2010quantum}%
  \BibitemOpen
  \bibfield  {author} {\bibinfo {author} {\bibfnamefont {M.~A.}\ \bibnamefont {Nielsen}}\ and\ \bibinfo {author} {\bibfnamefont {I.~L.}\ \bibnamefont {Chuang}},\ }\href@noop {} {\emph {\bibinfo {title} {Quantum computation and quantum information}}}\ (\bibinfo  {publisher} {Cambridge university press},\ \bibinfo {year} {2010})\BibitemShut {NoStop}%
\bibitem [{\citenamefont {Rieffel}\ and\ \citenamefont {Polak}(2011)}]{rieffel2011quantum}%
  \BibitemOpen
  \bibfield  {author} {\bibinfo {author} {\bibfnamefont {E.~G.}\ \bibnamefont {Rieffel}}\ and\ \bibinfo {author} {\bibfnamefont {W.~H.}\ \bibnamefont {Polak}},\ }\href@noop {} {\emph {\bibinfo {title} {Quantum computing: A gentle introduction}}}\ (\bibinfo  {publisher} {MIT press},\ \bibinfo {year} {2011})\BibitemShut {NoStop}%
\bibitem [{\citenamefont {DiVincenzo}\ and\ \citenamefont {Loss}(1999)}]{divincenzo1999quantum}%
  \BibitemOpen
  \bibfield  {author} {\bibinfo {author} {\bibfnamefont {D.~P.}\ \bibnamefont {DiVincenzo}}\ and\ \bibinfo {author} {\bibfnamefont {D.}~\bibnamefont {Loss}},\ }\bibfield  {title} {\enquote {\bibinfo {title} {Quantum computers and quantum coherence},}\ }\href@noop {} {\bibfield  {journal} {\bibinfo  {journal} {Journal of Magnetism and Magnetic Materials}\ }\textbf {\bibinfo {volume} {200}},\ \bibinfo {pages} {202} (\bibinfo {year} {1999})}\BibitemShut {NoStop}%
\bibitem [{\citenamefont {Knill}\ and\ \citenamefont {Laflamme}(1997)}]{knill1997theory}%
  \BibitemOpen
  \bibfield  {author} {\bibinfo {author} {\bibfnamefont {E.}~\bibnamefont {Knill}}\ and\ \bibinfo {author} {\bibfnamefont {R.}~\bibnamefont {Laflamme}},\ }\bibfield  {title} {\enquote {\bibinfo {title} {Theory of quantum error-correcting codes},}\ }\href@noop {} {\bibfield  {journal} {\bibinfo  {journal} {Physical Review A}\ }\textbf {\bibinfo {volume} {55}},\ \bibinfo {pages} {900} (\bibinfo {year} {1997})}\BibitemShut {NoStop}%
\bibitem [{\citenamefont {Lidar}\ \emph {et~al.}(1998)\citenamefont {Lidar}, \citenamefont {Chuang},\ and\ \citenamefont {Whaley}}]{lidar1998decoherence}%
  \BibitemOpen
  \bibfield  {author} {\bibinfo {author} {\bibfnamefont {D.~A.}\ \bibnamefont {Lidar}}, \bibinfo {author} {\bibfnamefont {I.~L.}\ \bibnamefont {Chuang}}, \ and\ \bibinfo {author} {\bibfnamefont {K.~B.}\ \bibnamefont {Whaley}},\ }\bibfield  {title} {\enquote {\bibinfo {title} {Decoherence-free subspaces for quantum computation},}\ }\href@noop {} {\bibfield  {journal} {\bibinfo  {journal} {Physical Review Letters}\ }\textbf {\bibinfo {volume} {81}},\ \bibinfo {pages} {2594} (\bibinfo {year} {1998})}\BibitemShut {NoStop}%
\bibitem [{\citenamefont {Zurek}(1991)}]{zurek1991decoherence}%
  \BibitemOpen
  \bibfield  {author} {\bibinfo {author} {\bibfnamefont {W.~H.}\ \bibnamefont {Zurek}},\ }\bibfield  {title} {\enquote {\bibinfo {title} {Decoherence and the transition from quantum to classical},}\ }\href@noop {} {\bibfield  {journal} {\bibinfo  {journal} {Physics today}\ }\textbf {\bibinfo {volume} {44}},\ \bibinfo {pages} {36} (\bibinfo {year} {1991})}\BibitemShut {NoStop}%
\bibitem [{\citenamefont {Schlosshauer}(2019)}]{schlosshauer2019quantum}%
  \BibitemOpen
  \bibfield  {author} {\bibinfo {author} {\bibfnamefont {M.}~\bibnamefont {Schlosshauer}},\ }\bibfield  {title} {\enquote {\bibinfo {title} {Quantum decoherence},}\ }\href@noop {} {\bibfield  {journal} {\bibinfo  {journal} {Physics Reports}\ }\textbf {\bibinfo {volume} {831}},\ \bibinfo {pages} {1} (\bibinfo {year} {2019})}\BibitemShut {NoStop}%
\bibitem [{\citenamefont {Chuang}\ \emph {et~al.}(1995)\citenamefont {Chuang}, \citenamefont {Laflamme}, \citenamefont {Shor},\ and\ \citenamefont {Zurek}}]{chuang1995quantum}%
  \BibitemOpen
  \bibfield  {author} {\bibinfo {author} {\bibfnamefont {I.~L.}\ \bibnamefont {Chuang}}, \bibinfo {author} {\bibfnamefont {R.}~\bibnamefont {Laflamme}}, \bibinfo {author} {\bibfnamefont {P.~W.}\ \bibnamefont {Shor}}, \ and\ \bibinfo {author} {\bibfnamefont {W.~H.}\ \bibnamefont {Zurek}},\ }\bibfield  {title} {\enquote {\bibinfo {title} {Quantum computers, factoring, and decoherence},}\ }\href@noop {} {\bibfield  {journal} {\bibinfo  {journal} {Science}\ }\textbf {\bibinfo {volume} {270}},\ \bibinfo {pages} {1633} (\bibinfo {year} {1995})}\BibitemShut {NoStop}%
\bibitem [{\citenamefont {Shor}(1996)}]{shor1996fault}%
  \BibitemOpen
  \bibfield  {author} {\bibinfo {author} {\bibfnamefont {P.~W.}\ \bibnamefont {Shor}},\ }\bibfield  {title} {\enquote {\bibinfo {title} {Fault-tolerant quantum computation},}\ }in\ \href@noop {} {\emph {\bibinfo {booktitle} {Proceedings of 37th conference on foundations of computer science}}}\ (\bibinfo {organization} {IEEE},\ \bibinfo {year} {1996})\ pp.\ \bibinfo {pages} {56--65}\BibitemShut {NoStop}%
\bibitem [{\citenamefont {Shor}(1995)}]{shor1995scheme}%
  \BibitemOpen
  \bibfield  {author} {\bibinfo {author} {\bibfnamefont {P.~W.}\ \bibnamefont {Shor}},\ }\bibfield  {title} {\enquote {\bibinfo {title} {Scheme for reducing decoherence in quantum computer memory},}\ }\href@noop {} {\bibfield  {journal} {\bibinfo  {journal} {Physical review A}\ }\textbf {\bibinfo {volume} {52}},\ \bibinfo {pages} {R2493} (\bibinfo {year} {1995})}\BibitemShut {NoStop}%
\bibitem [{\citenamefont {Kitaev}(2001)}]{kitaev2001unpaired}%
  \BibitemOpen
  \bibfield  {author} {\bibinfo {author} {\bibfnamefont {A.~Y.}\ \bibnamefont {Kitaev}},\ }\bibfield  {title} {\enquote {\bibinfo {title} {Unpaired majorana fermions in quantum wires},}\ }\href@noop {} {\bibfield  {journal} {\bibinfo  {journal} {Physics-uspekhi}\ }\textbf {\bibinfo {volume} {44}},\ \bibinfo {pages} {131} (\bibinfo {year} {2001})}\BibitemShut {NoStop}%
\bibitem [{\citenamefont {Leumer}\ \emph {et~al.}(2020)\citenamefont {Leumer}, \citenamefont {Marganska}, \citenamefont {Muralidharan},\ and\ \citenamefont {Grifoni}}]{leumer2020exact}%
  \BibitemOpen
  \bibfield  {author} {\bibinfo {author} {\bibfnamefont {N.}~\bibnamefont {Leumer}}, \bibinfo {author} {\bibfnamefont {M.}~\bibnamefont {Marganska}}, \bibinfo {author} {\bibfnamefont {B.}~\bibnamefont {Muralidharan}}, \ and\ \bibinfo {author} {\bibfnamefont {M.}~\bibnamefont {Grifoni}},\ }\bibfield  {title} {\enquote {\bibinfo {title} {Exact eigenvectors and eigenvalues of the finite kitaev chain and its topological properties},}\ }\href@noop {} {\bibfield  {journal} {\bibinfo  {journal} {Journal of Physics: condensed matter}\ }\textbf {\bibinfo {volume} {32}},\ \bibinfo {pages} {445502} (\bibinfo {year} {2020})}\BibitemShut {NoStop}%
\bibitem [{\citenamefont {Ran{\v{c}}i{\'c}}(2022)}]{ranvcic2022exactly}%
  \BibitemOpen
  \bibfield  {author} {\bibinfo {author} {\bibfnamefont {M.~J.}\ \bibnamefont {Ran{\v{c}}i{\'c}}},\ }\bibfield  {title} {\enquote {\bibinfo {title} {Exactly solving the kitaev chain and generating majorana-zero-modes out of noisy qubits},}\ }\href@noop {} {\bibfield  {journal} {\bibinfo  {journal} {Scientific Reports}\ }\textbf {\bibinfo {volume} {12}},\ \bibinfo {pages} {19882} (\bibinfo {year} {2022})}\BibitemShut {NoStop}%
\bibitem [{\citenamefont {Sung}\ \emph {et~al.}(2023)\citenamefont {Sung}, \citenamefont {Ran{\v{c}}i{\'c}}, \citenamefont {Lanes},\ and\ \citenamefont {Bronn}}]{sung2023simulating}%
  \BibitemOpen
  \bibfield  {author} {\bibinfo {author} {\bibfnamefont {K.~J.}\ \bibnamefont {Sung}}, \bibinfo {author} {\bibfnamefont {M.~J.}\ \bibnamefont {Ran{\v{c}}i{\'c}}}, \bibinfo {author} {\bibfnamefont {O.~T.}\ \bibnamefont {Lanes}}, \ and\ \bibinfo {author} {\bibfnamefont {N.~T.}\ \bibnamefont {Bronn}},\ }\bibfield  {title} {\enquote {\bibinfo {title} {Simulating majorana zero modes on a noisy quantum processor},}\ }\href@noop {} {\bibfield  {journal} {\bibinfo  {journal} {Quantum Science and Technology}\ }\textbf {\bibinfo {volume} {8}},\ \bibinfo {pages} {025010} (\bibinfo {year} {2023})}\BibitemShut {NoStop}%
\bibitem [{\citenamefont {Miao}\ \emph {et~al.}(2018)\citenamefont {Miao}, \citenamefont {Jin}, \citenamefont {Zhang},\ and\ \citenamefont {Zhou}}]{miao2018majorana}%
  \BibitemOpen
  \bibfield  {author} {\bibinfo {author} {\bibfnamefont {J.-J.}\ \bibnamefont {Miao}}, \bibinfo {author} {\bibfnamefont {H.-K.}\ \bibnamefont {Jin}}, \bibinfo {author} {\bibfnamefont {F.-C.}\ \bibnamefont {Zhang}}, \ and\ \bibinfo {author} {\bibfnamefont {Y.}~\bibnamefont {Zhou}},\ }\bibfield  {title} {\enquote {\bibinfo {title} {Majorana zero modes and long range edge correlation in interacting kitaev chains: analytic solutions and density-matrix-renormalization-group study},}\ }\href@noop {} {\bibfield  {journal} {\bibinfo  {journal} {Scientific reports}\ }\textbf {\bibinfo {volume} {8}},\ \bibinfo {pages} {488} (\bibinfo {year} {2018})}\BibitemShut {NoStop}%
\bibitem [{\citenamefont {Burnell}\ \emph {et~al.}(2013)\citenamefont {Burnell}, \citenamefont {Shnirman},\ and\ \citenamefont {Oreg}}]{burnell2013measuring}%
  \BibitemOpen
  \bibfield  {author} {\bibinfo {author} {\bibfnamefont {F.}~\bibnamefont {Burnell}}, \bibinfo {author} {\bibfnamefont {A.}~\bibnamefont {Shnirman}}, \ and\ \bibinfo {author} {\bibfnamefont {Y.}~\bibnamefont {Oreg}},\ }\bibfield  {title} {\enquote {\bibinfo {title} {Measuring fermion parity correlations and relaxation rates in one-dimensional topological superconducting wires},}\ }\href@noop {} {\bibfield  {journal} {\bibinfo  {journal} {Physical Review B}\ }\textbf {\bibinfo {volume} {88}},\ \bibinfo {pages} {224507} (\bibinfo {year} {2013})}\BibitemShut {NoStop}%
\bibitem [{\citenamefont {Fendley}(2012)}]{fendley2012parafermionic}%
  \BibitemOpen
  \bibfield  {author} {\bibinfo {author} {\bibfnamefont {P.}~\bibnamefont {Fendley}},\ }\bibfield  {title} {\enquote {\bibinfo {title} {Parafermionic edge zero modes in zn-invariant spin chains},}\ }\href@noop {} {\bibfield  {journal} {\bibinfo  {journal} {Journal of Statistical Mechanics: Theory and Experiment}\ }\textbf {\bibinfo {volume} {2012}},\ \bibinfo {pages} {P11020} (\bibinfo {year} {2012})}\BibitemShut {NoStop}%
\bibitem [{\citenamefont {Fendley}(2014)}]{fendley2014free}%
  \BibitemOpen
  \bibfield  {author} {\bibinfo {author} {\bibfnamefont {P.}~\bibnamefont {Fendley}},\ }\bibfield  {title} {\enquote {\bibinfo {title} {Free parafermions},}\ }\href@noop {} {\bibfield  {journal} {\bibinfo  {journal} {Journal of Physics A: Mathematical and Theoretical}\ }\textbf {\bibinfo {volume} {47}},\ \bibinfo {pages} {075001} (\bibinfo {year} {2014})}\BibitemShut {NoStop}%
\bibitem [{\citenamefont {Alicea}\ and\ \citenamefont {Fendley}(2016)}]{alicea2016topological}%
  \BibitemOpen
  \bibfield  {author} {\bibinfo {author} {\bibfnamefont {J.}~\bibnamefont {Alicea}}\ and\ \bibinfo {author} {\bibfnamefont {P.}~\bibnamefont {Fendley}},\ }\bibfield  {title} {\enquote {\bibinfo {title} {Topological phases with parafermions: theory and blueprints},}\ }\href@noop {} {\bibfield  {journal} {\bibinfo  {journal} {Annual Review of Condensed Matter Physics}\ }\textbf {\bibinfo {volume} {7}},\ \bibinfo {pages} {119} (\bibinfo {year} {2016})}\BibitemShut {NoStop}%
\bibitem [{\citenamefont {Jermyn}\ \emph {et~al.}(2014)\citenamefont {Jermyn}, \citenamefont {Mong}, \citenamefont {Alicea},\ and\ \citenamefont {Fendley}}]{jermyn2014stability}%
  \BibitemOpen
  \bibfield  {author} {\bibinfo {author} {\bibfnamefont {A.~S.}\ \bibnamefont {Jermyn}}, \bibinfo {author} {\bibfnamefont {R.~S.}\ \bibnamefont {Mong}}, \bibinfo {author} {\bibfnamefont {J.}~\bibnamefont {Alicea}}, \ and\ \bibinfo {author} {\bibfnamefont {P.}~\bibnamefont {Fendley}},\ }\bibfield  {title} {\enquote {\bibinfo {title} {Stability of zero modes in parafermion chains},}\ }\href@noop {} {\bibfield  {journal} {\bibinfo  {journal} {Physical Review B}\ }\textbf {\bibinfo {volume} {90}},\ \bibinfo {pages} {165106} (\bibinfo {year} {2014})}\BibitemShut {NoStop}%
\bibitem [{\citenamefont {Iemini}\ \emph {et~al.}(2017)\citenamefont {Iemini}, \citenamefont {Mora},\ and\ \citenamefont {Mazza}}]{iemini2017topological}%
  \BibitemOpen
  \bibfield  {author} {\bibinfo {author} {\bibfnamefont {F.}~\bibnamefont {Iemini}}, \bibinfo {author} {\bibfnamefont {C.}~\bibnamefont {Mora}}, \ and\ \bibinfo {author} {\bibfnamefont {L.}~\bibnamefont {Mazza}},\ }\bibfield  {title} {\enquote {\bibinfo {title} {Topological phases of parafermions: A model with exactly solvable ground states},}\ }\href@noop {} {\bibfield  {journal} {\bibinfo  {journal} {Physical Review Letters}\ }\textbf {\bibinfo {volume} {118}},\ \bibinfo {pages} {170402} (\bibinfo {year} {2017})}\BibitemShut {NoStop}%
\bibitem [{\citenamefont {Vaezi}(2013)}]{vaezi2013fractional}%
  \BibitemOpen
  \bibfield  {author} {\bibinfo {author} {\bibfnamefont {A.}~\bibnamefont {Vaezi}},\ }\bibfield  {title} {\enquote {\bibinfo {title} {Fractional topological superconductor with fractionalized majorana fermions},}\ }\href@noop {} {\bibfield  {journal} {\bibinfo  {journal} {Physical Review B}\ }\textbf {\bibinfo {volume} {87}},\ \bibinfo {pages} {035132} (\bibinfo {year} {2013})}\BibitemShut {NoStop}%
\bibitem [{\citenamefont {Vaezi}\ and\ \citenamefont {Vaezi}(2017{\natexlab{a}})}]{vaezi2017numerical}%
  \BibitemOpen
  \bibfield  {author} {\bibinfo {author} {\bibfnamefont {M.-S.}\ \bibnamefont {Vaezi}}\ and\ \bibinfo {author} {\bibfnamefont {A.}~\bibnamefont {Vaezi}},\ }\bibfield  {title} {\enquote {\bibinfo {title} {Numerical observation of parafermion zero modes and their stability in 2d topological states},}\ }\href@noop {} {\bibfield  {journal} {\bibinfo  {journal} {arXiv preprint arXiv:1706.01192}\ } (\bibinfo {year} {2017}{\natexlab{a}})}\BibitemShut {NoStop}%
\bibitem [{\citenamefont {Mahyaeh}(2020)}]{mahyaeh2020study}%
  \BibitemOpen
  \bibfield  {author} {\bibinfo {author} {\bibfnamefont {I.}~\bibnamefont {Mahyaeh}},\ }\emph {\bibinfo {title} {Study of the phase diagram of Zn symmetric chains}},\ \href@noop {} {Ph.D. thesis},\ \bibinfo  {school} {Department of Physics, Stockholm University} (\bibinfo {year} {2020})\BibitemShut {NoStop}%
\bibitem [{\citenamefont {Zhuang}\ \emph {et~al.}(2015)\citenamefont {Zhuang}, \citenamefont {Changlani}, \citenamefont {Tubman},\ and\ \citenamefont {Hughes}}]{zhuang2015phase}%
  \BibitemOpen
  \bibfield  {author} {\bibinfo {author} {\bibfnamefont {Y.}~\bibnamefont {Zhuang}}, \bibinfo {author} {\bibfnamefont {H.~J.}\ \bibnamefont {Changlani}}, \bibinfo {author} {\bibfnamefont {N.~M.}\ \bibnamefont {Tubman}}, \ and\ \bibinfo {author} {\bibfnamefont {T.~L.}\ \bibnamefont {Hughes}},\ }\bibfield  {title} {\enquote {\bibinfo {title} {Phase diagram of the z 3 parafermionic chain with chiral interactions},}\ }\href@noop {} {\bibfield  {journal} {\bibinfo  {journal} {Physical Review B}\ }\textbf {\bibinfo {volume} {92}},\ \bibinfo {pages} {035154} (\bibinfo {year} {2015})}\BibitemShut {NoStop}%
\bibitem [{\citenamefont {Wu}\ and\ \citenamefont {Lidar}(2002)}]{wu2002qubits}%
  \BibitemOpen
  \bibfield  {author} {\bibinfo {author} {\bibfnamefont {L.-A.}\ \bibnamefont {Wu}}\ and\ \bibinfo {author} {\bibfnamefont {D.}~\bibnamefont {Lidar}},\ }\bibfield  {title} {\enquote {\bibinfo {title} {Qubits as parafermions},}\ }\href@noop {} {\bibfield  {journal} {\bibinfo  {journal} {Journal of Mathematical Physics}\ }\textbf {\bibinfo {volume} {43}},\ \bibinfo {pages} {4506} (\bibinfo {year} {2002})}\BibitemShut {NoStop}%
\bibitem [{\citenamefont {Rao}(2017)}]{rao2017introduction}%
  \BibitemOpen
  \bibfield  {author} {\bibinfo {author} {\bibfnamefont {S.}~\bibnamefont {Rao}},\ }\bibfield  {title} {\enquote {\bibinfo {title} {Introduction to abelian and non-abelian anyons},}\ }\href@noop {} {\bibfield  {journal} {\bibinfo  {journal} {Topology and condensed matter physics}\ ,\ \bibinfo {pages} {399}} (\bibinfo {year} {2017})}\BibitemShut {NoStop}%
\bibitem [{\citenamefont {Narozhny}(2017)}]{narozhny2017majorana}%
  \BibitemOpen
  \bibfield  {author} {\bibinfo {author} {\bibfnamefont {B.}~\bibnamefont {Narozhny}},\ }\bibfield  {title} {\enquote {\bibinfo {title} {Majorana fermions in the nonuniform ising-kitaev chain: exact solution},}\ }\href@noop {} {\bibfield  {journal} {\bibinfo  {journal} {Scientific reports}\ }\textbf {\bibinfo {volume} {7}},\ \bibinfo {pages} {1447} (\bibinfo {year} {2017})}\BibitemShut {NoStop}%
\bibitem [{\citenamefont {Dasgupta}\ and\ \citenamefont {Ma}(1980)}]{dasgupta1980low}%
  \BibitemOpen
  \bibfield  {author} {\bibinfo {author} {\bibfnamefont {C.}~\bibnamefont {Dasgupta}}\ and\ \bibinfo {author} {\bibfnamefont {S.-k.}\ \bibnamefont {Ma}},\ }\bibfield  {title} {\enquote {\bibinfo {title} {Low-temperature properties of the random heisenberg antiferromagnetic chain},}\ }\href@noop {} {\bibfield  {journal} {\bibinfo  {journal} {Physical review b}\ }\textbf {\bibinfo {volume} {22}},\ \bibinfo {pages} {1305} (\bibinfo {year} {1980})}\BibitemShut {NoStop}%
\bibitem [{\citenamefont {Fisher}(1995)}]{fisher1995critical}%
  \BibitemOpen
  \bibfield  {author} {\bibinfo {author} {\bibfnamefont {D.~S.}\ \bibnamefont {Fisher}},\ }\bibfield  {title} {\enquote {\bibinfo {title} {Critical behavior of random transverse-field ising spin chains},}\ }\href@noop {} {\bibfield  {journal} {\bibinfo  {journal} {Physical review b}\ }\textbf {\bibinfo {volume} {51}},\ \bibinfo {pages} {6411} (\bibinfo {year} {1995})}\BibitemShut {NoStop}%
\bibitem [{\citenamefont {Shivamoggi}(2011)}]{shivamoggi2011majorana}%
  \BibitemOpen
  \bibfield  {author} {\bibinfo {author} {\bibfnamefont {V.~B.}\ \bibnamefont {Shivamoggi}},\ }\href@noop {} {\emph {\bibinfo {title} {Majorana fermions and Dirac edge states in topological phases}}}\ (\bibinfo  {publisher} {University of California, Berkeley},\ \bibinfo {year} {2011})\BibitemShut {NoStop}%
\bibitem [{\citenamefont {Benhemou}\ \emph {et~al.}(2023)\citenamefont {Benhemou}, \citenamefont {Angkhanawin}, \citenamefont {Adams}, \citenamefont {Browne},\ and\ \citenamefont {Pachos}}]{benhemou2023universality}%
  \BibitemOpen
  \bibfield  {author} {\bibinfo {author} {\bibfnamefont {A.}~\bibnamefont {Benhemou}}, \bibinfo {author} {\bibfnamefont {T.}~\bibnamefont {Angkhanawin}}, \bibinfo {author} {\bibfnamefont {C.~S.}\ \bibnamefont {Adams}}, \bibinfo {author} {\bibfnamefont {D.~E.}\ \bibnamefont {Browne}}, \ and\ \bibinfo {author} {\bibfnamefont {J.~K.}\ \bibnamefont {Pachos}},\ }\bibfield  {title} {\enquote {\bibinfo {title} {Universality of z 3 parafermions via edge-mode interaction and quantum simulation of topological space evolution with rydberg atoms},}\ }\href@noop {} {\bibfield  {journal} {\bibinfo  {journal} {Physical Review Research}\ }\textbf {\bibinfo {volume} {5}},\ \bibinfo {pages} {023076} (\bibinfo {year} {2023})}\BibitemShut {NoStop}%
\bibitem [{\citenamefont {White}(1992)}]{white1992density}%
  \BibitemOpen
  \bibfield  {author} {\bibinfo {author} {\bibfnamefont {S.~R.}\ \bibnamefont {White}},\ }\bibfield  {title} {\enquote {\bibinfo {title} {Density matrix formulation for quantum renormalization groups},}\ }\href@noop {} {\bibfield  {journal} {\bibinfo  {journal} {Physical review letters}\ }\textbf {\bibinfo {volume} {69}},\ \bibinfo {pages} {2863} (\bibinfo {year} {1992})}\BibitemShut {NoStop}%
\bibitem [{\citenamefont {White}(1993)}]{white1993density}%
  \BibitemOpen
  \bibfield  {author} {\bibinfo {author} {\bibfnamefont {S.~R.}\ \bibnamefont {White}},\ }\bibfield  {title} {\enquote {\bibinfo {title} {Density-matrix algorithms for quantum renormalization groups},}\ }\href@noop {} {\bibfield  {journal} {\bibinfo  {journal} {Physical review b}\ }\textbf {\bibinfo {volume} {48}},\ \bibinfo {pages} {10345} (\bibinfo {year} {1993})}\BibitemShut {NoStop}%
\bibitem [{\citenamefont {Schollw{\"o}ck}(2005)}]{schollwock2005density}%
  \BibitemOpen
  \bibfield  {author} {\bibinfo {author} {\bibfnamefont {U.}~\bibnamefont {Schollw{\"o}ck}},\ }\bibfield  {title} {\enquote {\bibinfo {title} {The density-matrix renormalization group},}\ }\href@noop {} {\bibfield  {journal} {\bibinfo  {journal} {Reviews of modern physics}\ }\textbf {\bibinfo {volume} {77}},\ \bibinfo {pages} {259} (\bibinfo {year} {2005})}\BibitemShut {NoStop}%
\bibitem [{\citenamefont {Schollw{\"o}ck}(2011)}]{schollwock2011density}%
  \BibitemOpen
  \bibfield  {author} {\bibinfo {author} {\bibfnamefont {U.}~\bibnamefont {Schollw{\"o}ck}},\ }\bibfield  {title} {\enquote {\bibinfo {title} {The density-matrix renormalization group in the age of matrix product states},}\ }\href@noop {} {\bibfield  {journal} {\bibinfo  {journal} {Annals of physics}\ }\textbf {\bibinfo {volume} {326}},\ \bibinfo {pages} {96} (\bibinfo {year} {2011})}\BibitemShut {NoStop}%
\bibitem [{\citenamefont {Verstraete}\ \emph {et~al.}(2023)\citenamefont {Verstraete}, \citenamefont {Nishino}, \citenamefont {Schollw{\"o}ck}, \citenamefont {Ba{\~n}uls}, \citenamefont {Chan},\ and\ \citenamefont {Stoudenmire}}]{verstraete2023density}%
  \BibitemOpen
  \bibfield  {author} {\bibinfo {author} {\bibfnamefont {F.}~\bibnamefont {Verstraete}}, \bibinfo {author} {\bibfnamefont {T.}~\bibnamefont {Nishino}}, \bibinfo {author} {\bibfnamefont {U.}~\bibnamefont {Schollw{\"o}ck}}, \bibinfo {author} {\bibfnamefont {M.~C.}\ \bibnamefont {Ba{\~n}uls}}, \bibinfo {author} {\bibfnamefont {G.~K.}\ \bibnamefont {Chan}}, \ and\ \bibinfo {author} {\bibfnamefont {M.~E.}\ \bibnamefont {Stoudenmire}},\ }\bibfield  {title} {\enquote {\bibinfo {title} {Density matrix renormalization group, 30 years on},}\ }\href@noop {} {\bibfield  {journal} {\bibinfo  {journal} {Nature Reviews Physics}\ }\textbf {\bibinfo {volume} {5}},\ \bibinfo {pages} {273} (\bibinfo {year} {2023})}\BibitemShut {NoStop}%
\bibitem [{\citenamefont {Vaezi}\ and\ \citenamefont {Vaezi}(2017{\natexlab{b}})}]{vaezi2017entanglement}%
  \BibitemOpen
  \bibfield  {author} {\bibinfo {author} {\bibfnamefont {M.-S.}\ \bibnamefont {Vaezi}}\ and\ \bibinfo {author} {\bibfnamefont {A.}~\bibnamefont {Vaezi}},\ }\bibfield  {title} {\enquote {\bibinfo {title} {Entanglement distance between quantum states and its implications for a density-matrix renormalization group study of degenerate ground states},}\ }\href@noop {} {\bibfield  {journal} {\bibinfo  {journal} {Physical Review B}\ }\textbf {\bibinfo {volume} {96}},\ \bibinfo {pages} {165129} (\bibinfo {year} {2017}{\natexlab{b}})}\BibitemShut {NoStop}%
\bibitem [{\citenamefont {Cobanera}\ and\ \citenamefont {Ortiz}(2014)}]{cobanera2014fock}%
  \BibitemOpen
  \bibfield  {author} {\bibinfo {author} {\bibfnamefont {E.}~\bibnamefont {Cobanera}}\ and\ \bibinfo {author} {\bibfnamefont {G.}~\bibnamefont {Ortiz}},\ }\bibfield  {title} {\enquote {\bibinfo {title} {Fock parafermions and self-dual representations of the braid group},}\ }\href@noop {} {\bibfield  {journal} {\bibinfo  {journal} {Physical Review A}\ }\textbf {\bibinfo {volume} {89}},\ \bibinfo {pages} {012328} (\bibinfo {year} {2014})}\BibitemShut {NoStop}%
\end{thebibliography}

%


\end{document}